\documentclass[aps,pra,reprint,amsmath,amssymb,superscriptaddress,longbibliography]{revtex4-2}

\usepackage[utf8]{inputenc}
\usepackage[T1]{fontenc}
\usepackage[spanish,english]{babel}
\usepackage{graphicx}
\usepackage{amsmath,amssymb,mathtools}
\usepackage{mathpazo} 
\usepackage{amsfonts}
\usepackage{amsthm}
\usepackage{bm}
\usepackage{mathrsfs}
\usepackage{physics}
\AtBeginDocument{}
\usepackage{xcolor}
\usepackage{hyperref}
\usepackage{orcidlink}
\usepackage[capitalize]{cleveref}
\usepackage{ragged2e}
\usepackage{float}
\usepackage{comment}

\newtheorem{prop}{Proposition}

\hypersetup{
  colorlinks=true,
  linkcolor=blue!70!black,
  citecolor=green!50!black,
  urlcolor=blue!80!black
}

\newcommand{\Id}{\mathbb{I}}
\newcommand{\Var}{\mathrm{Var}}

\DeclareMathOperator{\arctanh}{arctanh}
\newcommand{\ee}{\mathrm{e}}

\begin{document}

\title{Effective Thermodynamics of a Volume-Dependent Qubit}

\author{Andr\'{e} T. Ces\'{a}rio\,\orcidlink{0000-0002-6972-2576}}
\email{andretcs@ufmg.br}
\affiliation{Instituto de F\'{i}sica, Universidade Federal Fluminense, Niter\'{o}i, RJ, 24210-346, Brazil}
\affiliation{Departamento de F\'{i}sica, Universidade Federal de Pernambuco, Recife, PE, 50670-901, Brazil}

\author{Marcela Herrera\,\protect\orcidlink{https://orcid.org/0000-0001-7894-9206}}
\email{alba.herrera@correounivalle.edu.co }
\affiliation{Departamento de Física and Centre for Bioinformatics and Photonics—CIBioFi, Universidad del Valle, 760032 Cali, Colombia}

\begin{abstract}
\noindent We develop an effective thermodynamics for a qubit whose energy gap depends inversely on a volume-like control parameter, $\Delta(V)=\Delta_0/V$, where $V$ is a spectral-control coordinate rather than a literal mechanical volume. Starting from the canonical Gibbs state, we derive a complete thermodynamic structure from the Helmholtz potential and its Legendre and stability properties, including the fundamental relation $S(U,V)$, a generalized pressure conjugate to $V$, the equation of state $U=PV$, the first law, and the associated Maxwell relations. Reversible adiabats preserve the Gibbs populations and satisfy $TV=\mathrm{const.}$ and $PV^2=\mathrm{const.}$, corresponding to an effective exponent $\gamma=2$. The resulting effective Carnot cycle attains the Carnot efficiency, $\eta_{\mathrm{Carnot}}=1-T_C/T_H$, whereas the Otto cycle yields $\eta_{\mathrm{Otto}}=1-V_{\min}/V_{\max}$ and remains strictly sub-Carnot at finite work output. We identify the source of Otto irreversibility exactly: entropy is produced entirely during the isochoric thermalizations, and the cycle entropy production quantifies the informational mismatch between the adiabatically transported states and the reservoir-selected Gibbs states through the sum of the corresponding quantum relative entropies. Finally, we show that unitary-generated nonpassivity can act as a nonequilibrium battery, producing an apparent efficiency above Carnot when its preparation cost is omitted. At the same time, complete work accounting restores the thermodynamic bound.
\end{abstract}

\maketitle

\section{Introduction}
Thermodynamics was originally formulated for macroscopic systems, yet many of its most
basic notions--heat, work, entropy production, and reversibility--remain operationally meaningful when the working medium is reduced to a single
controllable quantum system. This observation has led to the development of quantum thermodynamics, where thermodynamic transformations are generated by
changes in the Hamiltonian, by thermalization with reservoirs, or by a combination of both \cite{Alicki1979,Kosloff2013,Goold2016,Vinjanampathy2016}. Quantum heat engines provide a natural framework for investigating these processes because they require a consistent separation between energy changes
caused by external control and energy changes caused by the redistribution of thermal populations \cite{Scovil1959,Kieu2004,Quan2007,Uzdin2015}. Recent
theoretical and experimental progress has shown that few-level systems, including single spins and superconducting artificial atoms, can operate as a genuine thermodynamic working medium whose performance is bounded by the same structural laws that constrain macroscopic engines
\cite{Peterson2019,Myers2022}.

A central issue in this context is the identification of thermodynamic control variables for microscopic systems. In a classical gas, the volume is a mechanical coordinate and the pressure is its conjugate force. In a finite quantum system, however, the experimentally controlled coordinate is often not a
geometric volume but a parameter that changes the spectrum of the Hamiltonian: a magnetic field in a spin system, a gate voltage in a quantum dot, a trapping frequency in an ion or cold-atom system, or a tunable coupling in a superconducting circuit. The thermodynamic meaning of such a control parameter
must therefore be derived from the equilibrium state and the corresponding thermodynamic potential, rather than imposed by analogy. Spin quantum heat
engines realized with nuclear magnetic resonance illustrate this point directly: the working medium is a spin-$1/2$ system, and work is associated
with controlled changes of the Zeeman gap rather than with mechanical displacement \cite{Peterson2019}. More generally, controllable two-level systems provide a minimal platform on which one can ask how a spectral control law generates an effective equation of state.

In this work, we address this question for a qubit whose energy gap depends on a volume-like parameter. The parameter $V$ is not assumed to be a literal macroscopic volume. Instead, it represents a spectral-control coordinate: increasing $V$ compresses the energy gap, while decreasing $V$ widens it. Such inverse scaling captures, at the level of a two-level effective model, the elementary fact that confinement energies decrease when an effective length scale increases. It also provides a simple bridge between volume-controlled thermodynamic cycles and experimentally relevant qubit platforms in which the level spacing is externally tunable~\cite{Koch2022QuantumOptimalControl,Oelsner2020TunableTwoLevelGap}.

The model is deliberately minimal. The Hamiltonian contains a single inverse gap-volume control law, and the equilibrium state is the corresponding canonical Gibbs state. From this starting point, the Helmholtz free energy defines entropy, internal energy, and a generalized pressure conjugate to the volume-like coordinate. Thus, the pressure is not introduced as an additional assumption; it emerges as the force associated with spectral control. The same construction reproduces the reversible first law from the microscopic energy balance of the qubit, connecting the thermodynamic and quantum-statistical descriptions without ambiguity.

This construction also makes it possible to compare standard thermodynamic cycles with their information-theoretic structure. For the volume-dependent qubit, reversible adiabats preserve the thermal population imbalance and therefore preserve the entropy. As a result, Carnot and Otto cycles can be described in terms of the same effective state variables, while their irreversibility has a direct interpretation in terms of mismatches between transported Gibbs populations and the Gibbs states imposed by the reservoirs. This point connects naturally with recent work in quantum thermodynamics showing that information, correlations, and coherence can modify the operational performance of quantum thermal machines~\cite{Uzdin2015,Goold2016,Micadei2019,Herrera2023}.

\begin{figure*}[t]
\centering
\includegraphics[width=0.92\textwidth]{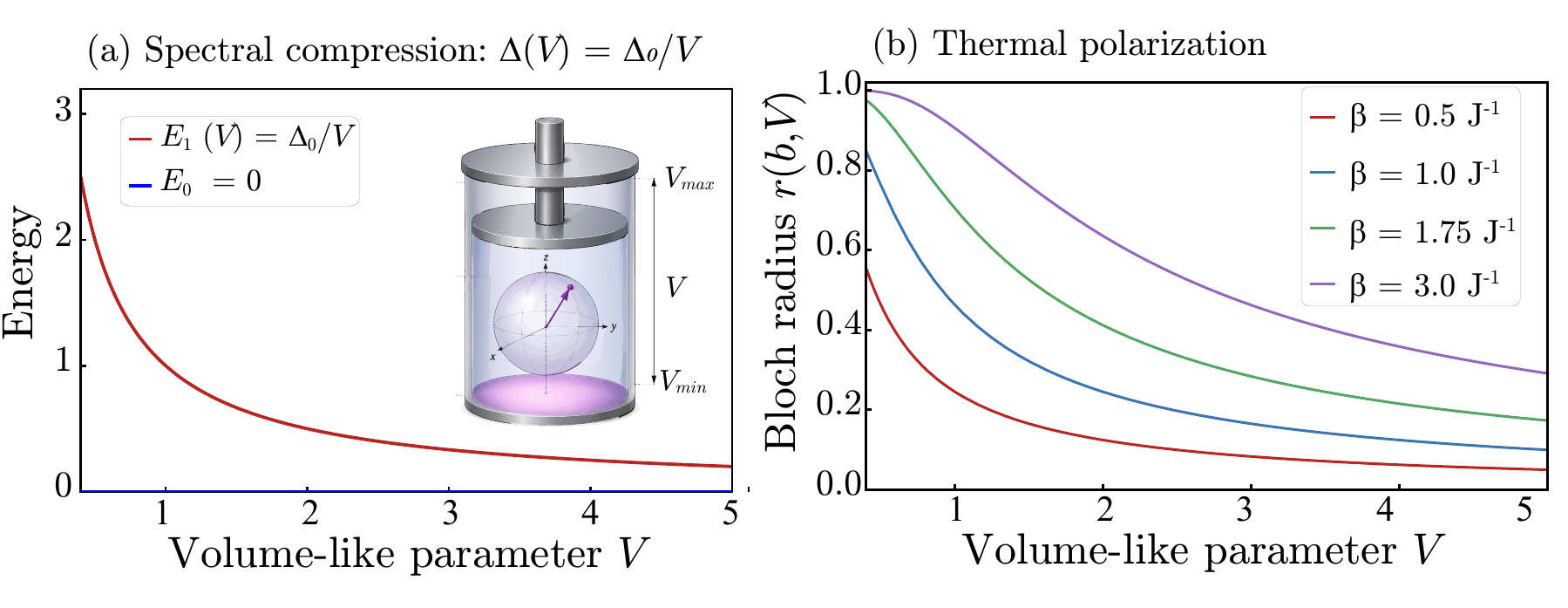}
\caption{The volume-dependent qubit model. (a) The excited-state energy $E_1(V)=\Delta_0/V$ [Eq.~\eqref{eq:H}] decreases monotonically as $V$ increases, while the ground state $E_0=0$ is unaffected. The inset offers a mechanical mnemonic for the abstract volume-like coordinate $V$. This piston-and-cavity picture is a visual aid only -- $V$ remains the spectral-control parameter of Eq.~\eqref{eq:H}, not a literal mechanical volume. (b) Bloch radius $r(\beta,V)=\tanh(\beta\Delta_0/2V)$ [Eq.~\eqref{eq:r}] as a function of $V$, for several values of the reservoir inverse temperature $\beta$: at fixed $\beta$, increasing $V$ closes the gap and drives the Gibbs state toward the maximally mixed state ($r\to0$), while decreasing $V$ polarizes it toward the ground projector $\Pi_0$ ($r\to1$). Both panels use $\Delta_0=1$.}
\label{fig:model}
\end{figure*}

Several related approaches have explored quantum heat engines based on single qubits, tunable energy gaps, and confined quantum working media. Single-qubit Otto and Carnot cycles have been analyzed from entropy-based and finite-time perspectives, addressing aspects such as coherence, quantum friction, or the operational cost of driving the system away from reversible behavior~\cite{Vallejo2025,Kosloff2013}. Other works consider particles in boxes or multilevel systems, where a geometric confinement length plays the role of a volume-like external parameter~\cite{Abe2013,Huang2018}. These studies provide a broader context for understanding how externally controlled energy spectra shape the microscopic analog of thermodynamic compression and expansion. The present work follows a complementary route: instead of starting from a mechanical box or from a generic externally driven qubit cycle, we construct an effective equilibrium thermodynamics directly from a volume-dependent two-level Hamiltonian. This allows us to define a generalized pressure, obtain an equation of state, and then analyze Carnot and Otto cycles within the same thermodynamic framework.

The paper is organized as follows. Section~\ref{sec:model} introduces the volume-dependent qubit Hamiltonian and derives its Gibbs state and Bloch representation. Section~\ref{sec:thermo} constructs the Helmholtz potential, the effective thermodynamic variables, the reversible first law, and the fundamental relation $S(U,V)$. Section~\ref{sec:response} analyzes the response functions and the associated thermodynamic geometry, and derives the reversible adiabatic curves. Sections~\ref{sec:carnot} and~\ref{sec:otto} construct the Carnot and Otto cycles, respectively, and Sec.~\ref{sec:info} discusses the information-theoretic interpretation of Otto irreversibility. Section~\ref{sec:conclusions} concludes. The Appendices~\ref{app:gibbs}--\ref{app:nonpassivity} collect the full algebraic derivations underlying the results quoted in the main text.

\section{Volume-dependent qubit and Gibbs state}
\label{sec:model}

We consider a two-level working medium with a Hamiltonian
\begin{equation}
H(V)=\Delta(V)\,\Pi_1,\qquad \Delta(V)=\frac{\Delta_0}{V},
\label{eq:H}
\end{equation}
where $[\Delta_0]=\mathrm{J}\,[V]$ and $V$ is expressed in the units of the spectral-control coordinate. $\Pi_0,\Pi_1$ are orthogonal projectors, $\Pi_0+\Pi_1=\mathbb{I}$. In the Bloch form, $\Pi_1=\tfrac12(\mathbb{I}+\hat n\cdot\vec\sigma)$, with $|\hat{n}| = 1$, so $H(V)=\tfrac{\Delta_0}{2V}(\mathbb{I}+\hat n\cdot\vec\sigma)$.

The parameter $V$ should not be viewed as a literal mechanical volume; it is a spectral-control coordinate. This inverse scaling reproduces, at the level of a two-level effective model, the elementary spectral compression seen, e.g., in a particle in a one-dimension box, $E_n\propto 1/L^2$, for $V\sim L^2$; more generally $\Delta(V)=\Delta_0 V^\alpha$ with $\alpha<0$ would serve the same purpose, and NMR Zeeman-gap control provides a direct experimental realization for the theory developed in this work~\cite{Peterson2019,Kieu2004,Oelsner2020TunableTwoLevelGap}. We adopt the inverse law throughout because it yields the most transparent effective thermodynamic structure; the qualitative construction below does not depend on this specific choice. Figure~\ref{fig:model} illustrates both effects together: panel (a) shows the spectral compression encoded in Eq.~\eqref{eq:H}, and panel (b) shows how this compression translates, through the canonical Gibbs state, into the thermal polarization discussed next.

The canonical Gibbs state $\rho(\beta,V)=e^{-\beta H(V)}/Z(\beta,V)$ has partition function
\begin{equation}
Z(\beta,V)=1+e^{-\beta\Delta(V)}=2e^{-x}\cosh x,
\label{eq:Z}
\end{equation}
where we introduce the dimensionless control variable
\begin{equation}
x:=\frac{\beta\Delta(V)}{2}=\frac{\beta\Delta_0}{2V}.
\label{eq:x}
\end{equation}
The Gibbs state takes the Bloch form
\begin{equation}
\rho(\beta,V)=\tfrac12\left[I-r(\beta,V)\,\hat n\cdot\vec\sigma\right],\qquad r:=p_0-p_1,
\label{eq:rho}
\end{equation}
with populations given by 
\begin{equation}
p_0=1/Z,\text{  and  } p_1=e^{-\beta\Delta(V)}/Z,
\label{eq:p_0_and_p_1}
\end{equation}
with Bloch radius defined by
\begin{equation}
r(\beta,V)=\tanh x = \tanh\!\left(\frac{\beta\Delta_0}{2V}\right).
\label{eq:r}
\end{equation}
For fixed $\beta$, decreasing $V$ increases the gap and polarizes the state toward the ground projector $\Pi_0$; increasing $V$ compresses the spectrum and drives $\rho$ toward the maximally mixed state (see Fig.~\ref{fig:model}). Thus $V$ controls the degree of thermal polarization at fixed temperature, and $(\beta,V)$ enters the Gibbs state only through the single combination $x$ of Eq.~\eqref{eq:x} -- a fact that will simplify the geometric and information-theoretic discussion throughout the paper. Appendix~\ref {app:gibbs} provides the explicit construction of the Gibbs state, its thermal populations, and its Bloch representation.

\section{Effective thermodynamics}
\label{sec:thermo}

\subsection{Helmholtz potential and thermodynamic variables}

From Eq.~\eqref{eq:Z}, the Helmholtz free energy $F=-\beta^{-1}\ln Z$ is
\begin{equation}
F(\beta,V)=\frac{\Delta_0}{2V}-\frac{1}{\beta}\ln\!\big(2\cosh x\big),
\label{eq:F}
\end{equation}
or, in terms of temperature, Eq.~\eqref{eq:F} can be written as $F(T,V)=\tfrac{\Delta_0}{2V}-k_BT\ln\!\big[2\cosh(\tfrac{\Delta_0}{2k_BTV})\big]$. We take $F$ as the fundamental potential and obtain entropy, pressure, and internal energy from the standard identities $S=-(\partial F/\partial T)_V$, $P=-(\partial F/\partial V)_T$, and $U=F+TS$, so that $P$ is not postulated, but emerges as the conjugate force to $V$. A direct computation, using $(\partial x/\partial\beta)_V=x/\beta$ and $(\partial x/\partial V)_\beta=-x/V$, gives
\begin{align}
S(x) &= k_B\big[\ln(2\cosh x)-x\tanh x\big], \label{eq:S}\\
P(V,x) &= \frac{\Delta_0}{2V^2}\big(1-\tanh x\big) = \frac{\Delta_0}{2V^2}(1-r), \label{eq:P}
\end{align}
with $U=F+TS=\tfrac{\Delta_0}{2V}(1-r)$. The same $U$ is obtained directly from the microscopic definition $U=\tr[\rho H]=\Delta(V)p_1$, for $p_1$ defined as in Eq.~\eqref{eq:p_0_and_p_1}, confirming the consistency of the two routes; we therefore only quote the result:
\begin{equation}
U(V,r)=\frac{\Delta_0}{2V}(1-r).
\label{eq:U}
\end{equation}
Equation~\eqref{eq:S} coincides with the von Neumann entropy $S(\rho) = -k_B\tr(\rho\ln\rho)$ of the binary distribution $\{p_0,p_1\}$, as can be checked by writing $p_{0,1}=e^{\pm x}/(2\cosh x)$.

Comparing Eqs.~\eqref{eq:P} and~\eqref{eq:U} gives the effective equation of state
\begin{equation}
U=PV,
\label{eq:eos}
\end{equation}
The effective equation of state introduced in Eq.~\eqref{eq:eos} is a model-dependent linear relation that follows directly from the spectral scaling law $\Delta(V)\propto V^{-1}$. Within this framework, the qubit admits an effective thermodynamic description formally analogous to the $U$--$P$--$V$ relation of a three-dimensional monatomic ideal gas. This analogy can be characterized through an effective compressibility factor, $z(x):=\frac{PV}{k_B T}=x\bigl(1-\tanh x\bigr)$, which quantifies deviations from the ideal-gas relation $PV=Nk_BT$ \cite{MoranShapiro,CengelBoles}. The compressibility factor is nonmonotonic and reaches a maximum at $x_*\simeq0.639232$. This point defines a crossover scale, rather than a phase transition, arising from the competition between the increase of the energy gap and the concomitant thermal depletion of the excited-state population as $V$ decreases.

The explicit derivation of the thermodynamic variables from the Helmholtz potential, including the equation of state and the crossover condition for $z(x)$, is given in Appendix~\ref{app:helmholtz}.

\subsection{First law and fundamental relation}

From $U=F+TS$ and $dF=-S\,dT-P\,\dd V$ one obtains the differential form of the first law
\begin{equation}
\dd U = T\,\dd S - P\,\dd V \equiv \delta Q - \delta W,
\label{eq:firstlaw}
\end{equation}
with $\delta Q=T\,\dd S$ and $\delta W=P\,\dd V$. This identification also follows directly from differentiating the microscopic energy $U(V,r)=\tfrac{\Delta_0}{2V}(1-r)$: writing $\dd U=(\partial_V U)_r\,\dd V+(\partial_r U)_V\,dr$ and using $\delta Q = T\,\dd S=-\tfrac{\Delta(V)}{2}dr$ [from Eq.~\eqref{eq:S} and $dr=\sech^2x\,\dd x$], one recovers $\delta W=P\,\dd V$ with the same $P$ as in Eq.~\eqref{eq:P}. Thus, heat is associated with changes of the thermal populations at fixed spectrum ($dr$), and work with changes of the spectrum at fixed populations ($\dd V$).

The microscopic heat--work decomposition and the derivation of the fundamental relation from the Gibbs populations are detailed in Appendix~\ref{app:firstlaw}.

Eliminating $\beta$ in favor of $U$ and $V$, with $y:=UV/\Delta_0=p_1$, the von Neumann entropy gives the fundamental relation
\begin{equation}
S(U,V) = -k_B\Big[(1-y)\ln(1-y)+y\ln y\Big],\quad y=\frac{UV}{\Delta_0},
\label{eq:SUV}
\end{equation}
valid on the positive-temperature domain $0\le UV/\Delta_0\le \tfrac12$, i.e. $0\le U\le \Delta_0/(2V)=E_1(V)/2$. Differentiating Eq.~\eqref{eq:SUV} reproduces $(\partial S/\partial U)_V=1/T$ and $(\partial S/\partial V)_U=P/T$ [using Eq.~\eqref{eq:eos}], so that $\dd S=\dd U/T+P\,\dd V/T$ is equivalent to Eq.~\eqref{eq:firstlaw}. The corresponding Maxwell relations, $(\partial(1/T)/\partial V)_U=(\partial(P/T)/\partial U)_V$ and $(\partial S/\partial V)_T=(\partial P/\partial T)_V$, hold identically and are verified explicitly in Appendix~\ref{app:firstlaw}. This closes the consistency check: the canonical free energy, the von Neumann entropy, the effective pressure, and $\langle H\rangle$ define a single and mutually consistent effective thermodynamic structure.

\section{Response functions and adiabats}
\label{sec:response}

The response functions have a general thermodynamic origin: the specific heat probes the response along the thermal direction, while the isothermal compressibility probes the response along the Hamiltonian-control direction $V$.
The heat capacity at fixed volume, $C_V=(\partial U/\partial T)_V$, follows from Eq.~\eqref{eq:U} as
\begin{equation}
C_V = k_B x^2\sech^2 x .
\label{eq:CV}
\end{equation}
This is the standard canonical fluctuation-response identity $C_V=k_B\beta^2\,\mathrm{Var}_\rho(H)$: since $H^2=\Delta(V)^2\Pi_1$, one has $\mathrm{Var}_\rho(H)=\Delta(V)^2p_1(1-p_1)=\tfrac14\Delta(V)^2\sech^2x$, which reproduces Eq.~\eqref{eq:CV} after multiplying by $k_B\beta^2$. Because $H(V)$ commutes with $\rho(\beta,V)$ for every $\beta,V$, this variance also coincides with the Bogoliubov--Kubo--Mori (BKM) covariance $g^{\rm BKM}_{\beta\beta}$~\cite{Petz1993,Petz1996,LesniewskiRuskai1999}, so $C_V=k_B\beta^2 g^{\rm BKM}_{\beta\beta}$: the specific heat measures the information-geometric susceptibility of the state under changes of $\beta$. Because $\rho(\beta,V)$ depends on $(\beta,V)$ only through $x$, the induced BKM line element $\dd\ell^2=\sech^2x\,\dd x^2$ has rank one when pulled back to the $(\beta,V)$ plane; its null directions, $\dd x=0$, are exactly the reversible adiabats discussed below. This foliation provides a geometric illustration of adiabatic inaccessibility within the equilibrium manifold, consistent with Carath\'eodory's formulation of the second law.

The isothermal compressibility $\kappa_T=-V^{-1}(\partial V/\partial P)_T$ is controlled by the curvature of $F$ along $V$,
\begin{equation}
\kappa_T=\frac{V^2}{\Delta_0\,\phi(x)},\qquad \phi(x):=1-\tanh x-\tfrac{x}{2}\sech^2x,
\label{eq:kappa}
\end{equation}
and the thermal expansion coefficient $\alpha_P=V^{-1}(\partial V/\partial T)_P$ is $\alpha_P= x\sech^2x/[2T\phi(x)]$. Both share the sign of $\phi(x)$, which vanishes at $x_\kappa\simeq1.10886$: for $x<x_\kappa$ the free-energy curvature is positive ($\kappa_T,\alpha_P>0$), and it changes sign above $x_\kappa$. Both $x_*$ and $x_\kappa$ are properties of the spectral law $\Delta(V)=\Delta_0/V$, not universal critical points; the cycles below are restricted to the regular branch $x<x_*<x_\kappa$. 

Detailed calculations of the heat capacity, the BKM pullback metric, the isothermal compressibility, and the thermal expansion coefficient are collected in Appendix~\ref{app:response}. Since $\dd S=-k_Bx\sech^2x\,\dd x$, a reversible adiabat ($\dd S=0$) satisfies $\dd x=0$, i.e.
\begin{equation}
x=\frac{\beta\Delta_0}{2V}=\text{const.}
\quad\Longleftrightarrow\quad
TV=\text{const.}
\label{eq:TV}
\end{equation}
This has the same form as the classical adiabatic law $TV^{\gamma-1}=\text{const.}$, immediately giving $\gamma=2$. Consistently, the Bloch radius $r=\tanh x$ remains constant along the adiabat, and from $P=\tfrac{\Delta_0}{2V^2}(1-r)$ one obtains
\begin{equation}
PV^{\gamma}=\text{const.},\qquad \gamma=2.
\label{eq:PV}
\end{equation}
Thus, the effective adiabatic exponent follows directly from $\dd S=0$ and the spectral law $\Delta(V)=\Delta_0/V$, rather than from equipartition. Its value is nevertheless consistent with the classical relation $\gamma=1+2/f$, for a classical ideal gas with $f=2$ active quadratic degrees of freedom. Here, however, $f$ should be understood only as the effective number of active degrees of freedom in this thermodynamic description. The reversible adiabatic relations, including the invariance of the Gibbs populations and the scaling laws $TV=\mathrm{const.}$ and $PV^2=\mathrm{const.}$, are derived in Appendix~\ref{app:adiabats}.

\section{Carnot engine}
\label{sec:carnot}

\begin{figure*}[t]
\centering
\includegraphics[width=1\textwidth]{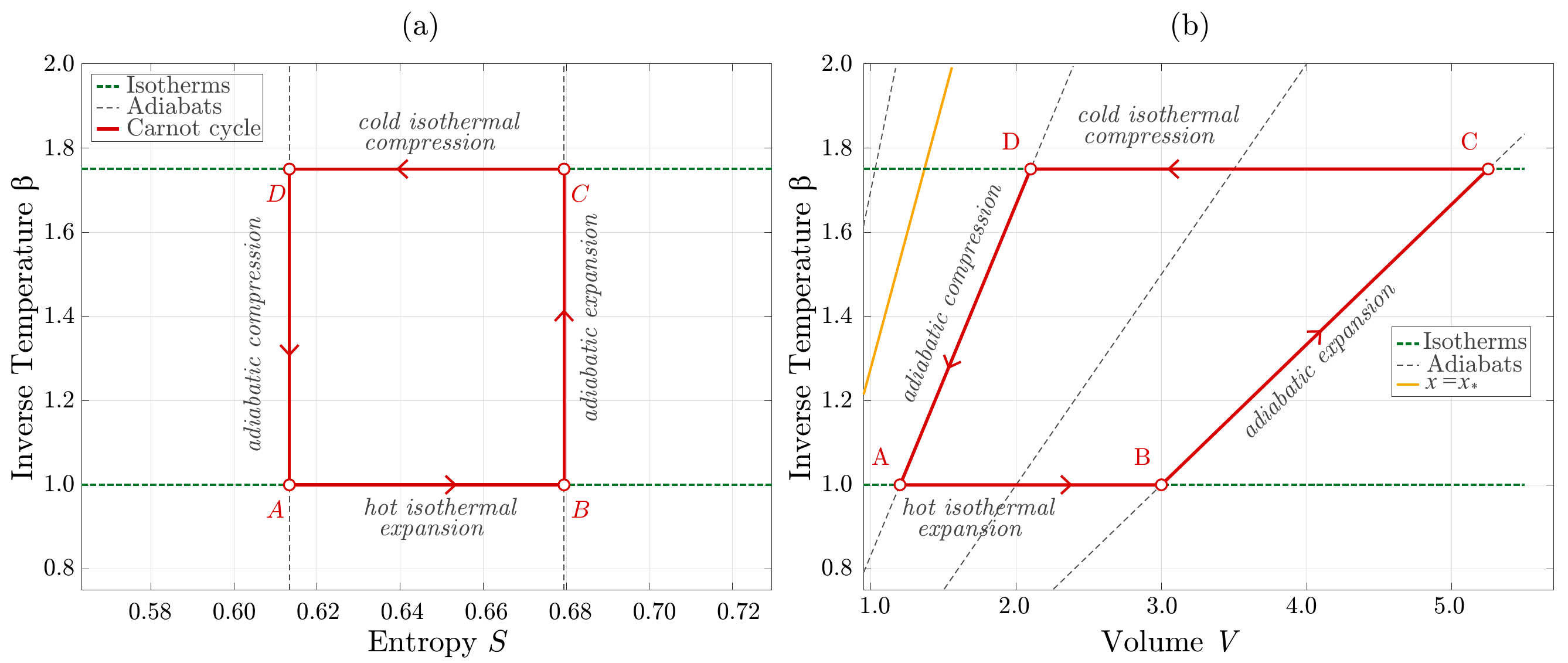}
\caption{\justifying Two planar representations of the Carnot cycle for the volume-dependent qubit. Panel (a): Thermodynamic representation in the $(\beta,S)$ plane, with entropy expressed in units of $k_B$. The isothermal branches are horizontal lines with $\beta=\beta_H$ and $\beta=\beta_C$, while the reversible adiabats are vertical lines with $S=\mathrm{const.}$ Panel (b): Control representation in the $(\beta,V)$ plane. The isothermal branches are again horizontal lines, whereas the reversible adiabats satisfy $x=\beta\Delta_0/(2V)=\mathrm{const.}$ The two diagrams describe the same reversible cycle: a hot isothermal expansion $A\to B$ at $\beta_H=1.0\mathrm{J^{-1}}$, a reversible adiabatic expansion $B\to C$ along $x=\mathrm{const.}$, a cold isothermal compression $C\to D$ at $\beta_C=1.75\mathrm{J^{-1}}$, and a reversible adiabatic compression $D\to A$. The cycle parameters are $V_A=1.20\mathrm{m^3}$, $V_B=3.00\mathrm{m^3}$, $V_C=5.25\mathrm{m^3}$, and $V_D=2.10\mathrm{m^3}$, satisfying $V_A<V_D<V_B<V_C$. The corresponding Carnot efficiency, for this example, is $\eta_{\mathrm{Carnot}}\approx 43\%$. The yellow line marks $x = x_*$, the maximum of the compressibility factor $z(x)$.}
\label{fig:carnot_planes}
\end{figure*}

\begin{figure*}[t]
\centering
\includegraphics[width=0.8\linewidth]{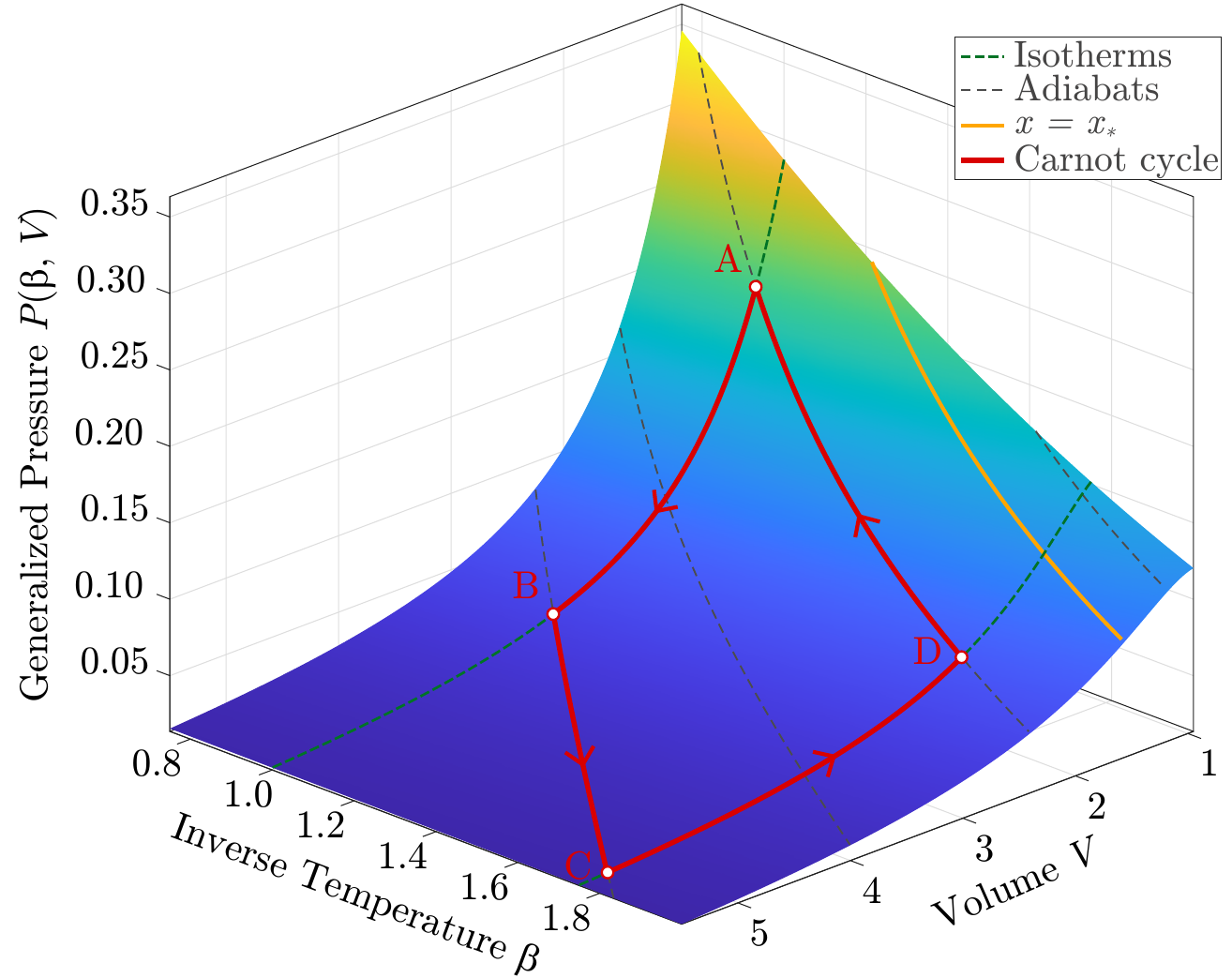}
\caption{\justifying Carnot cycle for the volume-dependent qubit represented on the pressure surface $P(\beta,V)$. The red curve is the same reversible cycle shown in Fig.~\ref{fig:carnot_planes}, now lifted to the effective $P-V-\beta$ diagram. The dotted curves are isotherms with $\beta=\mathrm{const.}$, while the dashed curves are reversible adiabats with $x=\beta\Delta_0/(2V)=\mathrm{const.}$ This reversible cycle comprises a hot isothermal expansion $A\to B$ at $\beta_H=1.0\mathrm{J^{-1}}$, a reversible adiabatic expansion $B\to C$ along $x=\mathrm{const.}$, a cold isothermal compression $C\to D$ at $\beta_C=1.75\mathrm{J^{-1}}$, and a reversible adiabatic compression $D\to A$. The cycle parameters are $V_A=1.20\mathrm{m^3}$, $V_B=3.00\mathrm{m^3}$, $V_C=5.25\mathrm{m^3}$, and $V_D=2.10\mathrm{m^3}$, satisfying $V_A<V_D<V_B<V_C$. The corresponding Carnot efficiency, for this example, is $\eta_{\mathrm{Carnot}}\approx 43\%$. The yellow line marks $x = x_*$, the maximum of the compressibility factor $z(x)$.}
\label{fig:carnot_surface}
\end{figure*}

We now use this structure to build a reversible Carnot cycle: a hot isothermal expansion $A\to B$ at $\beta_H$, a reversible adiabatic expansion $B\to C$, a cold isothermal compression $C\to D$ at $\beta_C>\beta_H$, and a reversible adiabatic compression $D\to A$, with corner volumes $V_A<V_D<V_B<V_C$ chosen so that $x_B=x_C$, $x_D=x_A$ [Eq.~\eqref{eq:TV} and \eqref{eq:PV}], and restricted to the regular regime $0<x_B<x_A<x_*<x_\kappa$. Figure~\ref{fig:carnot_planes} shows this cycle in two complementary planar representations, and Fig.~\ref{fig:carnot_surface} lifts it onto the effective pressure surface $P(\beta,V)$ of Eq.~\eqref{eq:P}.

For a generic isothermal branch at fixed $\beta_j^{\rm iso}$ between $x_i$ and $x_j$, integrating $\delta W=-\beta_j^{-1}(1-\tanh x)\dd x$ and $\delta Q=-\beta_j^{-1}x\sech^2x\,\dd x$ gives
\begin{align}
W_{i\to j} &= \frac{1}{\beta_j^{\rm iso}}\Big[x_i-x_j+\ln\!\Big(\frac{\cosh x_j}{\cosh x_i}\Big)\Big], \label{eq:Wiso}\\
Q_{i\to j} &= \frac{1}{\beta_j^{\rm iso}}\Big[x_i\tanh x_i - x_j\tanh x_j+\ln\!\Big(\frac{\cosh x_j}{\cosh x_i}\Big)\Big]. \label{eq:Qiso}
\end{align}
For a generic reversible adiabatic branch with $r_i=r_j=r_{\rm ad}$, $Q_{i\to j}=0$ and
\begin{equation}
W_{i\to j}=\frac{\Delta_0}{2}(1-r_{\rm ad})\Big(\frac{1}{V_i}-\frac{1}{V_j}\Big).
\label{eq:Wad}
\end{equation}

Applying Eq.~\eqref{eq:Wiso}--\eqref{eq:Qiso} to $A\to B$ (with $x_A>x_B$) gives $W_{AB}>0$ and $Q_{AB}>0$: the system absorbs heat and performs work, since $\cosh x_A/\cosh x_B<e^{x_A-x_B}$ guarantees the bracket in Eq.~\eqref{eq:Wiso} is positive. The internal energy varies as $\Delta U_{AB}=\beta_H^{-1}[z(x_B)-z(x_A)]<0$ because $z$ is increasing on $(0,x_*)$, so $W_{AB}>Q_{AB}$, the excess coming from the decrease of $U$. The adiabatic expansion $B\to C$ then delivers positive work at zero heat exchange [Eq.~\eqref{eq:Wad}], with $\Delta U_{BC}<0$. Along the cold isotherm $C\to D$ ($x_D=x_A>x_C=x_B$), Eqs.~\eqref{eq:Wiso}--\eqref{eq:Qiso} give $W_{CD}<0$ and $Q_{CD}<0$: work is done on the system and heat is released, with $\Delta U_{CD}=\beta_C^{-1}[z(x_A)-z(x_B)]>0$. Finally the adiabatic compression $D\to A$ closes the cycle with $Q_{DA}=0$, $W_{DA}<0$, $\Delta U_{DA}>0$.

Since the two adiabatic branches exchange no heat, the net work equals the net heat, $W_{\rm cyc}=Q_{AB}+Q_{CD}$; using Eqs.~\eqref{eq:Wiso}--\eqref{eq:Qiso} for both isotherms, one finds
\begin{equation}
W_{\rm cyc}=\Big(1-\frac{\beta_H}{\beta_C}\Big)Q_{AB}>0,
\end{equation}
and, with $Q_{\rm in}=Q_{AB}$,
\begin{equation}
\eta_{\rm Carnot}=\frac{W_{\rm cyc}}{Q_{\rm in}}=1-\frac{\beta_H}{\beta_C}=1-\frac{T_C}{T_H}.
\label{eq:etaCarnot}
\end{equation}
Thus, although the microscopic Hamiltonian fixes the shape of the cycle on the $(P,V,\beta)$ surface, the reversible Carnot efficiency depends only on the two reservoir temperatures, as it must. The complete branch-by-branch calculation of heat and work, together with the adiabatic constraints and the derivation of the net cycle efficiency, is presented in Appendix~\ref{app:carnot}.

\section{Otto engine}
\label{sec:otto}

\begin{figure*}[t]
\centering
\includegraphics[width=1\linewidth]{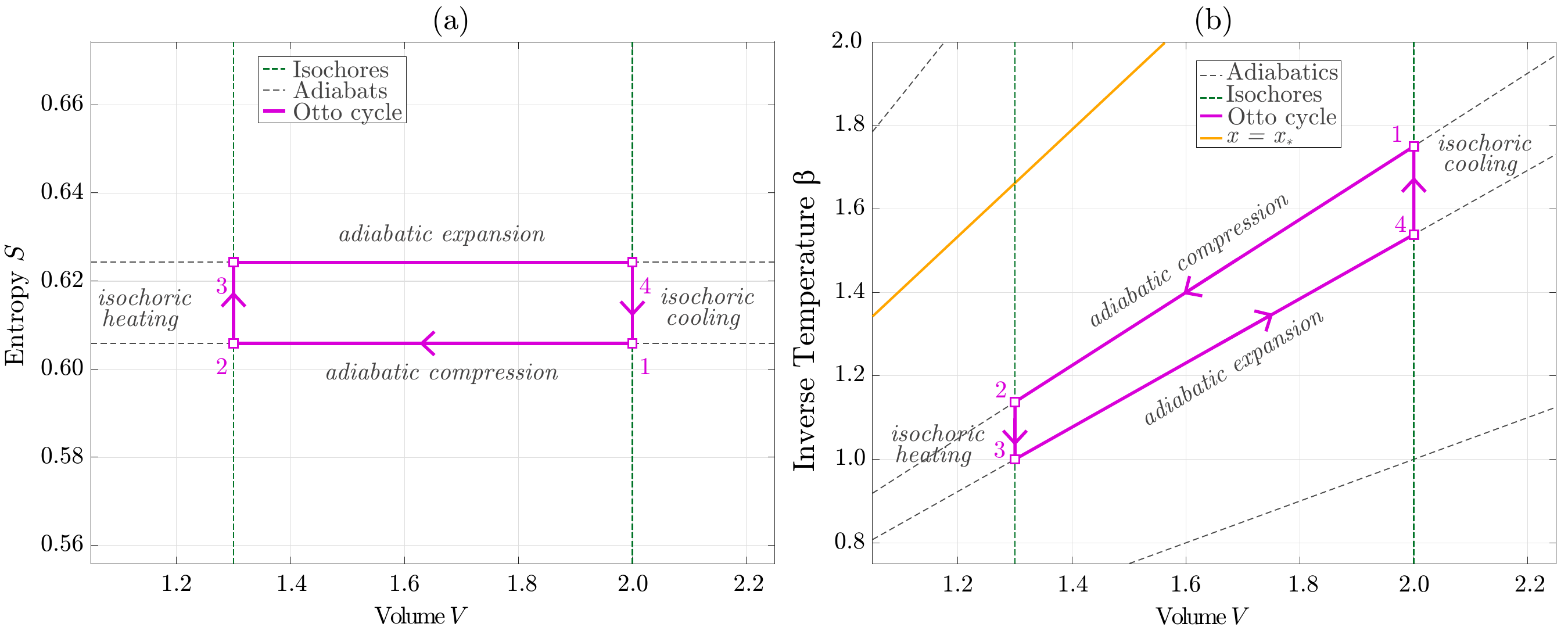}
\caption{
Two planar representations of the Otto cycle for the volume-dependent qubit. Panel (a): Representation in the $(V,S)$ plane, where the reversible adiabatic branches are horizontal, with $S_1=S_2$ and $S_3=S_4$, and the isochoric branches are vertical, with $V_2=V_3=V_{\min}$ and $V_1=V_4=V_{\max}$. Entropy is shown in units of $k_B$. Panel (b): Representation in the $(V,\beta)$ plane. In this representation, the isochoric branches remain at fixed $V$, whereas the reversible adiabatic branches satisfy $x=\beta\Delta_0/(2V)=\mathrm{const.}$ The four strokes are the reversible adiabatic compression $1\to2$, the isochoric heating $2\to3$, the reversible adiabatic expansion $3\to4$, and the isochoric cooling $4\to1$. The reservoir inverse temperatures are assigned to points $1$ and $3$, with $1=(\beta_C,V_{\max})$ and $3=(\beta_H,V_{\min})$, while the intermediate inverse temperatures are associated with points $2$ and $4$, namely $2=(\beta_2,V_{\min})$ and $4=(\beta_4,V_{\max})$, with $\beta_2$ and $\beta_4$ fixed by the adiabatic conditions. The plotted cycle uses representative parameters $V_{\min}=V_2=V_3=1.3\,\mathrm{m^3}$, $V_{\max}=V_1=V_4=2.0\,\mathrm{m^3}$, $\beta_H=1.0\,\mathrm{J^{-1}}$, $\beta_C=1.75\,\mathrm{J^{-1}}$, $\beta_2=1.1375\,\mathrm{J^{-1}}$, and $\beta_4\simeq1.5385\,\mathrm{J^{-1}}$.
 The yellow line marks $x = x_*$, the maximum of the compressibility factor $z(x)$.}
\label{fig:otto_planes}
\end{figure*}

\begin{figure*}[t]
\centering
\includegraphics[width=0.8\linewidth]{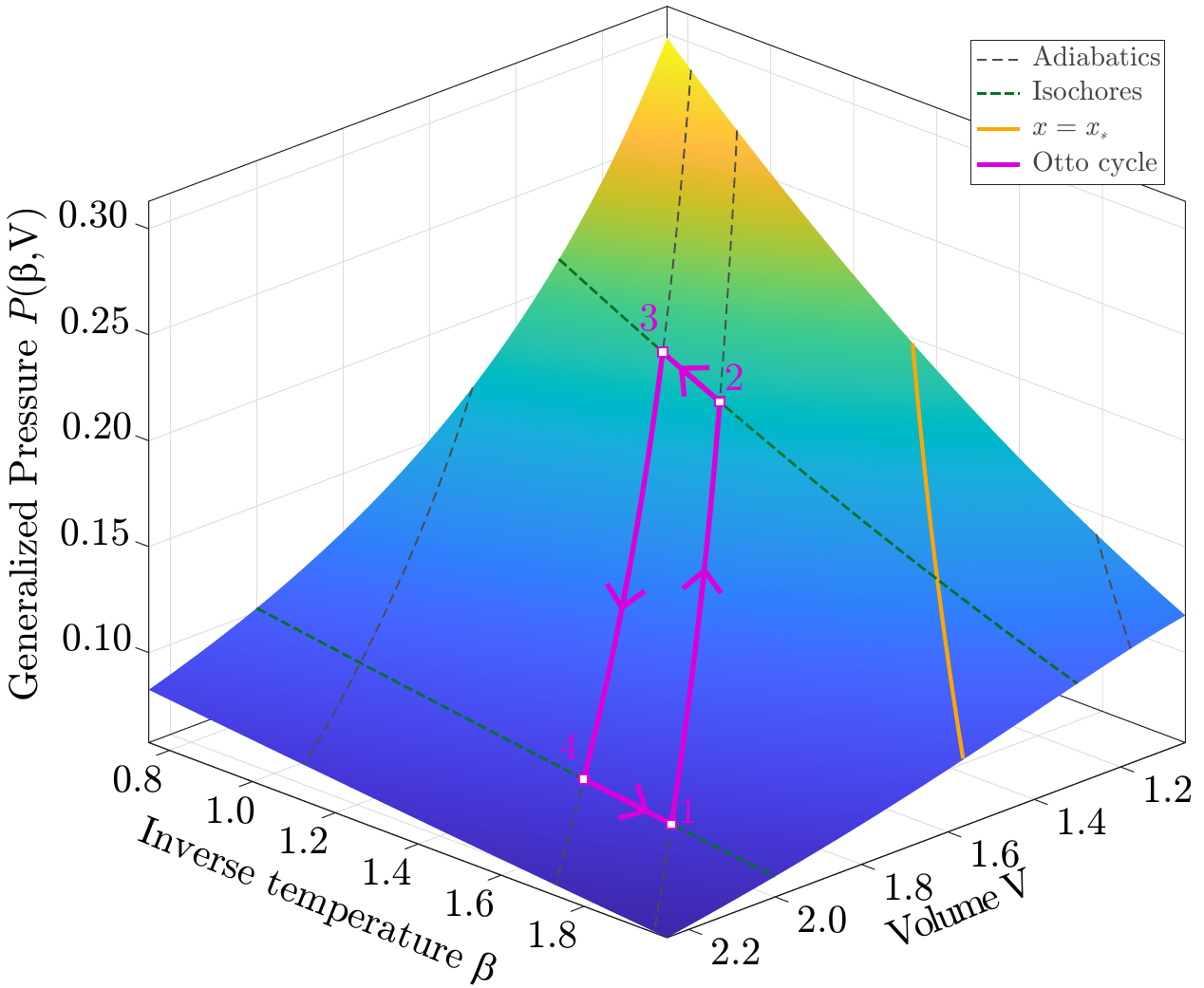}
\caption{
The Otto cycle on the generalized-pressure surface
$P(\beta,V)$ of the volume-dependent qubit. The magenta curve is the Otto cycle, the dotted curves are isochores with
$V=\mathrm{const.}$, and the dashed curves are reversible adiabats with $x=\beta\Delta_0/(2V)=\mathrm{const.}$ The corner states are $\rho_1=\rho(\beta_C,V_{\max})$,
$\rho_2=\rho(\beta_2,V_{\min})$,
$\rho_3=\rho(\beta_H,V_{\min})$, and
$\rho_4=\rho(\beta_4,V_{\max})$, where $\beta_2$ and $\beta_4$ are the effective inverse temperatures generated by the
population-preserving adiabatic branches. The branch $1\to2$ is a
reversible adiabatic compression, $2\to3$ is isochoric heating,
$3\to4$ is a reversible adiabatic expansion, and $4\to1$ is
isochoric cooling.  The yellow line marks $x = x_*$, the maximum of the compressibility factor $z(x)$.}
\label{fig:otto_surface}
\end{figure*}

The Otto cycle consists of a reversible adiabatic compression $1\to2$ ($V_{\max}\to V_{\min}$), isochoric heating $2\to3$ at $V_{\min}$, a reversible adiabatic expansion $3\to4$ ($V_{\min}\to V_{\max}$), and isochoric cooling $4\to1$ at $V_{\max}$. The adiabatic branches preserve $x$ and $r$: $x_1=x_2$, $x_3=x_4$; the reservoir temperatures $\beta_C,\beta_H$ are assigned to points $1,3$, while $\beta_2=\beta_C V_{\min}/V_{\max}$ and $\beta_4=\beta_H V_{\max}/V_{\min}$ follow from the adiabatic condition, Eq.~\eqref{eq:TV} and \eqref{eq:PV}. Figure~\ref{fig:otto_planes} shows the cycle in the $(V,S)$ and $(V,\beta)$ planes, and Fig.~\ref{fig:otto_surface} lifts it onto the pressure surface; the relative-entropy mismatches of Sec.~\ref{sec:info} that quantify its irreversibility are associated with the two isochoric strokes $2\to3$ and $4\to1$.

On an isochoric branch at fixed $V_{\rm iso}$, $\delta W=0$ and, from Eq.~\eqref{eq:U}, $Q_{i\to j}=\Delta U_{i\to j}=\tfrac{\Delta_0}{2V_{\rm iso}}(r_i-r_j)$, so the sign of the exchanged heat is set by the change of Bloch radius alone. Using Eq.~\eqref{eq:Wad} for the two adiabats and this isochoric formula for the two strokes at $V_{\min},V_{\max}$, the net cycle work reduces to the two adiabatic contributions,
\begin{equation}
W_{\rm Otto}=\frac{\Delta_0}{2}(r_1-r_3)\Big(\frac{1}{V_{\min}}-\frac{1}{V_{\max}}\Big).
\end{equation}
The cycle produces positive work and absorbs heat from the hot reservoir along $2\to3$, provided that
\begin{equation}
r_1=r_2>r_3=r_4 \iff x_1>x_3 \iff \frac{V_{\max}}{V_{\min}}<\frac{\beta_C}{\beta_H}.
\label{eq:ottocondition}
\end{equation}
This is the Otto engine condition: at fixed reservoir temperatures, the expansion ratio cannot be arbitrarily large if the cycle is to produce net work. With $Q_{\rm in}=Q_{23}=\tfrac{\Delta_0}{2V_{\min}}(r_1-r_3)$, the Otto efficiency is
\begin{equation}
\eta_{\rm Otto}=\frac{W_{\rm Otto}}{Q_{\rm in}}=1-\frac{V_{\min}}{V_{\max}}=1-\frac{\epsilon_{\min}}{\epsilon_{\max}},
\label{eq:etaOtto}
\end{equation}
with $\epsilon_{\min,\max}=\Delta_0/V_{\max,\min}$ the two extremal gaps. This is the standard Otto efficiency for spectra that scale uniformly with a control parameter~\cite{Kosloff2013,Quan2007}; what is new here is that the same result emerges from the derived equation of state and from the relative-entropy structure of its deficit. Since Eq.~\eqref{eq:ottocondition} implies $V_{\min}/V_{\max}>T_C/T_H$, one has $\eta_{\rm Otto}<\eta_{\rm Carnot}$, as expected. Eq.~\eqref{eq:etaOtto} coincides with the classical ideal-gas Otto efficiency $1-R^{1-\gamma}$ evaluated at the effective exponent $\gamma=2$ derived in Eq.~\eqref{eq:TV}--\eqref{eq:PV}, where $R=V_{\max}/V_{\min}$; here, however, $\gamma=2$ is not a phenomenological heat-capacity ratio but a consequence of the Hamiltonian control law $\Delta(V)=\Delta_0/V$ and the equation of state $U=PV$. The individual heat and work contributions of the four Otto strokes, the positive-work condition, and the resulting efficiency are derived explicitly in Appendix~\ref{app:otto}.

\section{Information-theoretic interpretation}
\label{sec:info}

Throughout this section, entropies are measured in units of Boltzmann's constant, such that $S(\rho)=-\Tr(\rho\ln\rho)$, and all entropy productions $\Sigma$ are dimensionless. The corresponding physical entropy production is $k_B\Sigma$.

\subsection{Thermalization mismatch}

The reversible adiabatic strokes of the Otto cycle preserve the Gibbs parameter $x=\beta\Delta_0/(2V)$ and therefore preserve both the Bloch radius $r=\tanh x$ and the entropy. In particular, $x_1=x_2$, $x_3=x_4$, $r_1=r_2$, and $r_3=r_4$. These strokes transport the Gibbs populations to a different Hamiltonian spectrum without producing entropy. Consequently, the state entering each isochoric stroke remains Gibbsian with respect to the Hamiltonian reached after the preceding adiabat and to the transported inverse temperature, $\beta_2$ or $\beta_4$. It is not, however, the Gibbs state selected by the reservoir that subsequently acts. The relevant nonequilibrium condition is therefore not a loss of Gibbs form, but a thermalization mismatch between two Gibbs states associated with the same Hamiltonian and different inverse temperatures.

The irreversibility of the Otto cycle is thus entirely associated with the two isochoric thermalization strokes. Since the working medium returns to its initial state after one complete cycle, its entropy change vanishes, and the total dimensionless entropy production is given by the reservoir entropy balance,
\begin{equation}
\Sigma_{\mathrm{Otto}}
=
\beta_C Q_{\mathrm{out}}
-
\beta_H Q_{\mathrm{in}},
\label{eq:Sigma_Otto_reservoirs_final}
\end{equation}
where $Q_{\mathrm{in}}=Q_{23}>0$ and $Q_{\mathrm{out}}=-Q_{41}>0$. Using $Q_{\mathrm{in}}=\Delta_0(r_1-r_3)/(2V_{\min})$ and $Q_{\mathrm{out}}=\Delta_0(r_1-r_3)/(2V_{\max})$, together with $x_1=\beta_C\Delta_0/(2V_{\max})$ and $x_3=\beta_H\Delta_0/(2V_{\min})$, Eq.~\eqref{eq:Sigma_Otto_reservoirs_final} becomes
\begin{equation}
\mathcal M_{\mathrm{Otto}}
=
\Sigma_{\mathrm{Otto}}
=
(x_1-x_3)(r_1-r_3),
\label{eq:M_Sigma_Otto}
\end{equation}
where $\mathcal M_{\mathrm{Otto}}$ denotes the total thermalization mismatch of the cycle. For the complete isochoric thermalizations considered here, the fixed-Hamiltonian relative-entropy identity derived in Appendix~\ref{app:mismatch} shows that this mismatch is exactly equal to the entropy produced by the cycle. Whenever no ambiguity arises, we use the shorthand $\mathcal M\equiv\mathcal M_{\mathrm{Otto}}$, which is also the notation adopted in Figure~\ref{fig:otto_surface}.

The factorized form in Eq.~\eqref{eq:M_Sigma_Otto} resembles the bilinear structure commonly encountered in irreversible thermodynamics, where entropy production is expressed through conjugate thermodynamic mismatches \cite{Onsager1931I,Onsager1931II,DeGrootMazur}. In the present model, this structure follows directly from $\dd s=-x\,dr$, which identifies $x$ and $r$ as conjugate variables for the Gibbs family. The difference $x_1-x_3$ quantifies the thermal mismatch between the reservoir-selected equilibrium states, whereas $r_1-r_3$ is the corresponding population mismatch transported by the adiabatic strokes. Equivalently, $\beta_CQ_{\mathrm{out}}=x_1(r_1-r_3)$ and $\beta_HQ_{\mathrm{in}}=x_3(r_1-r_3)$: the two reservoirs act on the same transported population interval but weight it with different values of the conjugate parameter $x$. Since the adiabats satisfy $dr=0$, they do not contribute to the entropy production.

The finite-volume origin of the mismatch follows from the adiabatic transport of the reservoir temperatures. Defining the compression ratio as $R=V_{\max}/V_{\min}$, conservation of $x$ along the adiabats gives $\beta_2=\beta_C/R$ after the compression $1\to2$ and $\beta_4=\beta_HR$ after the expansion $3\to4$. The hot isochore must therefore remove the mismatch $\beta_2-\beta_H$, while the cold isochore removes $\beta_C-\beta_4$. In terms of the Gibbs parameter, this gives
\begin{equation}
x_1-x_3
=
\frac{\Delta_0}{2V_{\min}}
\left(
\frac{\beta_C}{R}-\beta_H
\right)
=
\frac{\Delta_0}{2V_{\max}}
\left(
\beta_C-\beta_HR
\right).
\label{eq:finite_volume_mismatch}
\end{equation}
Substitution of Eq.~\eqref{eq:finite_volume_mismatch} into Eq.~\eqref{eq:M_Sigma_Otto} yields the finite-volume representation
\begin{equation}
\mathcal M_{\mathrm{Otto}}
=
\frac{\Delta_0}{2V_{\min}}
\left(
\frac{\beta_C}{R}-\beta_H
\right)
(r_1-r_3).
\label{eq:finite_volume_M}
\end{equation}

Equation~\eqref{eq:finite_volume_M} makes the origin of the irreversibility transparent. The engine regime requires $1<R<\beta_C/\beta_H=T_H/T_C$, which is equivalent to $\beta_2>\beta_H$ and $\beta_4<\beta_C$. Consequently, $x_1>x_3$ and $r_1>r_3$, so that $\mathcal M_{\mathrm{Otto}}>0$ for every nontrivial engine cycle.

The mismatch vanishes only as $R\to\beta_C/\beta_H$, when the transported temperatures become exactly matched to the reservoirs, $\beta_2\to\beta_H$ and $\beta_4\to\beta_C$. In the same limit, $x_1\to x_3$, $r_1-r_3\to0$, and therefore both $Q_{\mathrm{in}}$ and $W_{\mathrm{Otto}}$ vanish. Thus, although $\eta_{\mathrm{Otto}}=1-1/R$ approaches the Carnot value in this limit, it does so only in the degenerate zero-work regime. Any Otto engine with finite heat absorption and finite work output necessarily operates with a nonzero thermalization mismatch.

The relative-entropy interpretation of this result is developed in Appendix~\ref{app:mismatch}. There, the standard fixed-Hamiltonian relation between entropy production and the decrease of relative entropy to the reservoir-selected Gibbs state is applied separately to the two isochores. This relation is a specialization of the connection between dissipation and statistical distinguishability in stochastic thermodynamics \cite{Kawai2007,EspositoVanDenBroeck2010,Seifert2012,Parrondo2015}, and of its quantum formulation based on the monotonicity of quantum relative entropy and Spohn's entropy-production inequality \cite{Lindblad1975,Spohn1978,DeffnerLutz2010,Goold2016}. For the present Otto cycle, the two isochoric contributions add to
$\Sigma_{\mathrm{Otto}}=D(\rho_2\Vert\rho_3)+D(\rho_4\Vert\rho_1)$.

From Eq.~\eqref{eq:Sigma_Otto_reservoirs_final}, one directly obtains $\Sigma_{\mathrm{Otto}}=\beta_CQ_{\mathrm{in}}(\eta_{\mathrm{Carnot}}-\eta_{\mathrm{Otto}})$. Using the relative-entropy representation derived in Appendix~\ref{app:mismatch}, the corresponding efficiency deficit becomes
\begin{equation}
\eta_{\mathrm{Otto}}
=
\eta_{\mathrm{Carnot}}
-
\frac{
D(\rho_2\Vert\rho_3)
+
D(\rho_4\Vert\rho_1)
}{
\beta_CQ_{\mathrm{in}}
}.
\label{eq:eta_otto_relative_entropy}
\end{equation}
Equation~\eqref{eq:eta_otto_relative_entropy} identifies the reduction of the Otto efficiency relative to the Carnot benchmark directly with the two thermalization mismatches removed during the hot and cold isochores.

The same relation can be written directly in terms of the work output. Defining $W_{\mathrm{Otto}}=\eta_{\mathrm{Otto}}Q_{\mathrm{in}}$ and $W_{\mathrm{Carnot}}=\eta_{\mathrm{Carnot}}Q_{\mathrm{in}}$, we obtain
\begin{equation}
\Sigma_{\mathrm{Otto}}
=
\beta_C
\left(
W_{\mathrm{Carnot}}
-
W_{\mathrm{Otto}}
\right).
\label{eq:Sigma_work_deficit}
\end{equation}
Equivalently,
\begin{equation}
W_{\mathrm{Otto}}
=
W_{\mathrm{Carnot}}
-
\frac{\Sigma_{\mathrm{Otto}}}{\beta_C}.
\label{eq:Otto_work_deficit}
\end{equation}
Equations~\eqref{eq:Sigma_work_deficit} and \eqref{eq:Otto_work_deficit} show that the entropy produced during the two isochoric thermalizations determines the work deficit relative to the reversible Carnot benchmark at the same absorbed heat. Since $\mathcal M_{\mathrm{Otto}}=\Sigma_{\mathrm{Otto}}$, the same deficit is quantified by the thermalization mismatch used throughout the figures. In the reduced description, the information displaced from the working medium during thermalization is no longer accessible as a work resource to the engine.

\subsection{Unitary-generated nonpassivity as a nonequilibrium battery}

Thermalization mismatch characterizes a loss mechanism: nonequilibrium free energy that reaches a reservoir without being extracted is dissipated during relaxation. A nonequilibrium state may, however, contain useful free energy that can be converted into work before thermalization. In the present qubit model, the relevant resource is not coherence alone, but the nonpassivity generated by a coherence-creating unitary.

Consider the Gibbs state $\rho_{\mathrm{eq}}(\beta,V)=p_0\Pi_0+p_1\Pi_1$, with $r=p_0-p_1=\tanh[\beta\Delta(V)/2]$. Maximization over its unitary orbit, whose elementary derivation is given in  Appendix~\ref{app:nonpassivity}, gives $|c|_{\max}=r/2$. A representative maximally coherent state is
\begin{equation}
\rho_{\mathrm{coh}}^{\max}
=
{\small\frac12}\begin{pmatrix}
1 & r\\
r & 1
\end{pmatrix}.
\label{eq:rho_coh_max}
\end{equation}
Equation~\eqref{eq:rho_coh_max} preserves the spectrum and entropy of the original Gibbs state but changes its energy-basis populations. The corresponding thermodynamic resource is therefore the nonpassivity generated by the coherence-creating unitary, rather than coherence alone.

The preparation is performed at fixed Hamiltonian and in isolation from the reservoirs, so that $Q_{\mathrm{aux}}=0$. The equilibrium and maximally coherent energies are $U_{\mathrm{eq}}=\Delta(V)(1-r)/2$ and $U_{\mathrm{coh}}^{\max}=\Delta(V)/2$, respectively. The external drive therefore stores the nonequilibrium free energy
$\Delta F_{\mathrm{np}}=\Delta U_{\mathrm{np}}=\Delta(V)r/2$. In this operational sense, the working medium acts as a nonequilibrium battery charged by an external work source.

A meaningful comparison with the unassisted Otto cycle requires the auxiliary unitary protocol to form a closed charge--discharge loop before the subsequent thermal contact. Denoting the charging and discharging unitaries by $\mathcal V_{\mathrm{ch}}$ and $\mathcal V_{\mathrm{dis}}$, respectively, this condition is
\begin{equation}
\rho_{\mathrm{coh}}^{\max}
=
\mathcal V_{\mathrm{ch}}
\rho_{\mathrm{eq}}
\mathcal V_{\mathrm{ch}}^\dagger,
\label{eq:auxiliary_closed_loop}
\end{equation}
and for the discharge operation $\rho_{\mathrm{eq}} = \mathcal V_{\mathrm{dis}}
\rho_{\mathrm{coh}}^{\max}
\mathcal V_{\mathrm{dis}}^\dagger.$
The closure condition in Eq.~\eqref{eq:auxiliary_closed_loop} ensures that the assisted and unassisted cycles reach the next thermal stroke with identical boundary state and Hamiltonian. Consequently, their subsequent thermal trajectory, $Q_{\mathrm{in}}$, and $\mathcal M_{\mathrm{Otto}}$ are the same. If the discharge is incomplete, residual populations or coherences reach the reservoir, and the assisted device is a different thermodynamic cycle, for which the heat exchanges and thermalization mismatch must be recomputed. The relation between the work recovered during discharge and the mismatch left for the subsequent thermal reset is summarized by the following proposition.

\begin{prop}[Ergotropic extraction and thermal reset of unitary-generated nonpassivity]
\label{prop:ergotropic_reset}
Let $H(V)=\Delta(V)\Pi_1$, with $\Delta(V)=\Delta_0/V$, and let $\rho_{\mathrm{eq}}(\beta,V)=p_0\Pi_0+p_1\Pi_1$, with $p_0\ge p_1$, be the corresponding Gibbs state. For the maximally coherent state in Eq.~\eqref{eq:rho_coh_max}, the ergotropy \cite{Allahverdyan2004} is
\begin{equation}
\mathcal W_{\max}
:=
\mathcal W
\left(
\rho_{\mathrm{coh}}^{\max},H(V)
\right)
=
\frac{\Delta(V)}{2}
\tanh\!\left(
\frac{\beta\Delta(V)}{2}
\right).
\label{eq:ergotropy_max}
\end{equation}
If the state is completely thermalized without prior work extraction, the entropy produced is
\begin{equation}
\Sigma_{\mathrm{np}}^{\mathrm{dir}}
=
D\!\left(
\rho_{\mathrm{coh}}^{\max}
\Vert
\rho_{\mathrm{eq}}
\right)
=
\beta\mathcal W_{\max}.
\label{eq:direct_reset_ergotropy}
\end{equation}
More generally, consider a unitary discharge
$\rho_{\mathrm d}=\mathcal V_{\mathrm d}\rho_{\mathrm{coh}}^{\max}\mathcal V_{\mathrm d}^{\dagger}$ extracting
$W_{\mathrm d}=\Tr[H(V)(\rho_{\mathrm{coh}}^{\max}-\rho_{\mathrm d})]$, with $0\le W_{\mathrm d}\le\mathcal W_{\max}$. A subsequent complete thermal reset produces
\begin{equation}
\Sigma_{\mathrm{res}}(\mathcal V_{\mathrm d})
=
D\!\left(
\rho_{\mathrm d}
\Vert
\rho_{\mathrm{eq}}
\right)
=
\beta
\left(
\mathcal W_{\max}-W_{\mathrm d}
\right).
\label{eq:residual_mismatch}
\end{equation}
Equivalently,
\begin{equation}
\beta W_{\mathrm d}
+
\Sigma_{\mathrm{res}}(\mathcal V_{\mathrm d})
=
\beta\mathcal W_{\max}.
\label{eq:ergotropy_mismatch_balance}
\end{equation}
With the local Gibbs parameter $x_{\beta,V}=\beta\Delta(V)/2=\beta\Delta_0/(2V)$, the right-hand side of Eq.~\eqref{eq:ergotropy_mismatch_balance} is $x_{\beta,V}\tanh x_{\beta,V}$.
\end{prop}

Equation~\eqref{eq:direct_reset_ergotropy} describes direct dissipation of the entire unitary-generated work resource, whereas Eq.~\eqref{eq:residual_mismatch} describes the mismatch remaining after an arbitrary unitary discharge. Their common balance, Eq.~\eqref{eq:ergotropy_mismatch_balance}, separates two limiting uses of the same nonequilibrium resource. If $W_{\mathrm d}=0$, the entire nonpassive work value reaches the reservoir and appears as thermalization mismatch. If $W_{\mathrm d}=\mathcal W_{\max}$, optimal extraction returns the system to its passive Gibbs representative and no residual mismatch remains. A partial discharge interpolates continuously between these limits: one part of the resource is recovered as useful work, while the remainder is displaced from the working medium during the subsequent thermal reset. The charging cost, auxiliary cycle closure, and residual mismatch after partial discharge are derived in Appendix~\ref{app:nonpassivity}.

Under the closed-loop condition in Eq.~\eqref{eq:auxiliary_closed_loop}, an ideal discharge can be compared with the original Otto cycle without modifying its heat input. If the external charging cost is excluded from the input bookkeeping, the apparent thermal efficiency is
\begin{equation}
\eta_{\mathrm{th,app}}^{\max}
=
\eta_{\mathrm{Carnot}}
+
\frac{
\beta_C\mathcal W_{\max}
-
\mathcal M_{\mathrm{Otto}}
}{
\beta_CQ_{\mathrm{in}}
}.
\label{eq:apparent_efficiency}
\end{equation}
Equation~\eqref{eq:apparent_efficiency} shows that the apparent thermal efficiency exceeds the Carnot value whenever $\beta_C\mathcal W_{\max}>\mathcal M_{\mathrm{Otto}}$. The excess work does not originate from the hot reservoir; it is recovered from the externally charged nonequilibrium battery. A related bookkeeping effect occurs in correlation-assisted quantum engines, where an apparent enhancement beyond the standard Carnot value can arise when pre-existing correlations are consumed without being included among the thermodynamic resources \cite{Herrera2023}.

Once the charging cost is included, $\mathcal W_{\max}=\Delta F_{\mathrm{np}}$ implies that the ideal auxiliary charge--discharge loop contributes zero net work. The complete heat-engine efficiency is therefore
\begin{equation}
\eta_{\mathrm{net}}
=
\eta_{\mathrm{Otto}}
=
\eta_{\mathrm{Carnot}}
-
\frac{
\mathcal M_{\mathrm{Otto}}
}{
\beta_CQ_{\mathrm{in}}
}
\le
\eta_{\mathrm{Carnot}}.
\label{eq:net_efficiency_nonpassive}
\end{equation}
Equation~\eqref{eq:net_efficiency_nonpassive} shows that the unitary-generated battery can modify the gross work output when its charging cost is kept outside the thermal bookkeeping, but it does not alter the Carnot bound for the complete thermodynamic process.

Finally, the preceding mismatch balance refers to the reduced, ideal-reservoir description. In a microscopic treatment with an explicit finite reservoir $R$, information removed from the working medium is not destroyed but redistributed between the reservoir and system--reservoir correlations. For a product reference state, the relative-entropy chain rule gives
\begin{equation}
D\!\left(
\rho_{SR}'
\Vert
\rho_{\mathrm{eq}}\otimes\rho_E
\right) =
D\!\left(
\rho_S'
\Vert
\rho_{\mathrm{eq}}
\right)
+
D\!\left(
\rho_E'
\Vert
\rho_E
\right) +
I(S\!:\!E)_{\rho'}.
\label{eq:relative_entropy_chain_rule}
\end{equation}
Here $I(S\!:\!E)_{\rho'}=S(\rho_S')+S(\rho_E')-S(\rho_{SE}')\ge0$ is the mutual information generated between system and reservoir. Equation~\eqref{eq:relative_entropy_chain_rule} shows explicitly that reservoir perturbations and system--reservoir correlations provide additional nonnegative contributions to the information balance beyond the mismatch retained in the reduced working medium, as in finite-reservoir formulations of Landauer's principle \cite{ReebWolf2014}. The losses associated with thermalization should therefore be understood as a loss of accessible work from the working medium, rather than as information destruction.

\section{Conclusions}
\label{sec:conclusions}

We have shown that a two-level system with an inverse gap-volume control law, $\Delta(V)=\Delta_0/V$, supports a complete and self-consistent effective equilibrium thermodynamics: the Helmholtz free energy generated by the canonical Gibbs state produces entropy, internal energy, and a generalized pressure conjugate to $V$ that satisfy the reversible first law and the equation of state $U=PV$, with no assumption beyond the spectral control law itself. Reversible adiabats preserve the Bloch radius and obey $PV^2=\text{const.}$, giving an effective adiabatic exponent $\gamma=2$, as discussed in Eq.~\eqref{eq:PV}.

Within this framework, the Carnot cycle built from the qubit saturates the usual Carnot efficiency $1-T_C/T_H$, while the Otto cycle is strictly sub-Carnot, with an efficiency deficit that we derived exactly: it equals the sum of two quantum relative entropies quantifying the population mismatch between the state delivered by each reversible adiabat and the Gibbs state subsequently imposed by the corresponding reservoir. This gives a fully analytic, minimal-model illustration of the general principle that thermodynamic irreversibility in cyclic quantum engines is controlled by distinguishability from reservoir-selected equilibrium states, and it does so with a control parameter, $V$, whose thermodynamic conjugate force is derived rather than assumed.

Because the construction depends on the Hamiltonian only through the ratio $x=\beta\Delta_0/(2V)$, the results here can be adapted directly to other spectral control laws $\Delta(V)\propto V^\alpha$ with $\alpha<0$, which would modify the effective adiabatic exponent and the details of the response functions, but not the overall consistency of the thermodynamic construction or the relative-entropy origin of Otto irreversibility. 

A particularly relevant extension concerns multipartite and many-body working media, where interactions and correlations can become genuine thermodynamic resources. In such systems, the generalized pressure and response functions would acquire correlation-dependent contributions. At the same time, the Gibbs manifold would no longer be restricted to the effectively one-dimensional geometry of the present qubit model. This opens the possibility of studying how classical and quantum correlations modify the effective equation of state, thermodynamic susceptibilities, and the information-theoretic cost of irreversibility. In the many-body limit, the same framework could further clarify how collective thermodynamic behavior, extensivity, and genuine critical phenomena emerge from microscopic spectral control. Natural extensions include finite-time and finite-reservoir corrections to the mismatch identity in Eq.~\eqref{eq:M_Sigma_Otto}, following finite-size Landauer bounds~\cite{ReebWolf2014}, and experimental implementations of the volume-like control law in NMR, gate-tunable quantum dots, or superconducting qubits with tunable couplings~\cite{Peterson2019,Oelsner2020TunableTwoLevelGap}.

\begin{acknowledgments}
ATC acknowledges CNPq through process No.~150643/2026-8, and also RAU No.~12-2016-AY-UNA, Per\'{u}.
\end{acknowledgments}

\bibliographystyle{unsrt}
\bibliography{ref}

\hypertarget{sec:appendix}{}
\twocolumngrid
\appendix
\setcounter{theorem}{0}


\phantomsection 
\section*{Appendices}
\label{appendices}
The appendices collect the algebraic derivations underlying the results quoted in the main text. We use the convention that work performed by the working medium is positive, so that the first law is $\dd U=\delta Q-\delta W$. Thermodynamic entropies carry units of $k_B$ unless we explicitly introduce the dimensionless entropy $s=S/k_B$.

\section{Gibbs state, Bloch radius, and the volume-dependent qubit}
\label{app:gibbs}

The working medium is a two-level system with Hamiltonian
\begin{equation}
H(V)=\Delta(V)\Pi_1,
\qquad
\Delta(V)=\frac{\Delta_0}{V},
\label{app:H}
\end{equation}
where $\Pi_0+\Pi_1=\Id$, $\Pi_i\Pi_j=\delta_{ij}\Pi_i$, and the energy eigenvalues are
\begin{equation}
E_0(V)=0,
\qquad
E_1(V)=\Delta(V)=\frac{\Delta_0}{V}.
\end{equation}
Using the Bloch representation,
\begin{equation}
\Pi_0=\frac12(\Id-\hat n\cdot\bm\sigma),
\quad
\Pi_1=\frac12(\Id+\hat n\cdot\bm\sigma),
\quad |\hat n|=1,
\end{equation}
the Hamiltonian can be written as
\begin{equation}
H(V)
=
\frac{\Delta_0}{2V}
\left(
\mathbb{I}
+
\hat n\cdot\vec{\sigma}
\right).
\end{equation}
The canonical state at inverse temperature $\beta=(k_BT)^{-1}$ is
\begin{equation}
\rho(\beta,V)=\frac{\ee^{-\beta H(V)}}{Z(\beta,V)}.
\end{equation}
Because $H(V)=0\,\Pi_0+\Delta(V)\Pi_1$,
\begin{equation}
\ee^{-\beta H(V)}=\Pi_0+\ee^{-\beta\Delta(V)}\Pi_1,
\end{equation}
and therefore, the partition function is
\begin{equation}
Z(\beta,V)=1+\ee^{-\beta\Delta(V)}.
\label{app:Zraw}
\end{equation}
The thermal populations are consequently
\begin{equation}
p_0=\frac{1}{1+\ee^{-\beta\Delta}},
\quad
p_1=\frac{\ee^{-\beta\Delta}}{1+\ee^{-\beta\Delta}},
\quad
p_0+p_1=1.
\label{app:pops}
\end{equation}

It is convenient to introduce
\begin{equation}
x:=\frac{\beta\Delta(V)}{2}=\frac{\beta\Delta_0}{2V}.
\label{app:xdef}
\end{equation}
Then
\begin{equation}
Z=1+\ee^{-2x}=2\ee^{-x}\cosh x,
\label{app:Zx}
\end{equation}
and
\begin{equation}
p_0=\frac{\ee^x}{2\cosh x},
\qquad
p_1=\frac{\ee^{-x}}{2\cosh x}.
\label{app:px}
\end{equation}
For $p_0>p_1$, the population imbalance defines the Bloch radius $r$:
\begin{equation}
r:=p_0-p_1,
\end{equation}
it follows that
\begin{align}
r
&=\frac{1-\ee^{-2x}}{1+\ee^{-2x}}
=\frac{\ee^x-\ee^{-x}}{\ee^x+\ee^{-x}}
=\tanh x.
\label{app:r}
\end{align}
Thus, the dimensionless variable $x$ and the volume-like control coordinate $V(\beta,r)$ are
\begin{equation}
x=\arctanh r,
\qquad
V(\beta,r)=\frac{\beta\Delta_0}{2\arctanh r}.
\label{app:Vbr}
\end{equation}
The Gibbs state is 
\begin{align}
\rho
&=p_0\Pi_0+p_1\Pi_1
=\frac12\left[\Id-r\,\hat n\cdot\bm\sigma\right].
\label{app:rhoBloch}
\end{align}
Thus, the Gibbs family depends on the pair $(\beta,V)$ only through the single dimensionless combination $x=\beta\Delta_0/(2V)$.

\section{Helmholtz potential and effective thermodynamic variables}
\label{app:helmholtz}

From Eq.~\eqref{app:Zx},
\begin{align}
F(\beta,V)
&=-\frac{1}{\beta}\ln Z
=-\frac{1}{\beta}\left[-x+\ln(2\cosh x)\right],
\nonumber\\
&=\frac{\Delta_0}{2V}-\frac{1}{\beta}\ln\left[2\cosh\left(\frac{\beta\Delta_0}{2V}\right)\right].
\label{app:F}
\end{align}
Equivalently,
\begin{equation}
F(T,V)=\frac{\Delta_0}{2V}
-k_BT\ln\!\left[2\cosh\!\left(\frac{\Delta_0}{2k_BTV}\right)\right].
\end{equation}

\subsection{Internal energy from the microscopic definition}

Using Eqs.~\eqref{app:H} and \eqref{app:pops}, and since $p_1=(1-r)/2$,
\begin{equation}
U(V,r) =\Tr(\rho H)
=\frac{\Delta_0}{V}p_1=\frac{\Delta_0}{2V}(1-r).
\label{app:U}
\end{equation}
Substituting $r=\tanh x$ and $\Delta_0/(2V)=x/\beta$,
\begin{equation}
U(\beta,x)=\frac{x}{\beta}(1-\tanh x).
\label{app:Ux}
\end{equation}

\subsection{Entropy from the Helmholtz potential}

We now derive the entropy and the effective pressure from the Helmholtz free energy. At fixed $V$,
$\left(\frac{\partial x}{\partial\beta}\right)_V
=\frac{\Delta_0}{2V}=\frac{x}{\beta}$, and, because
\begin{equation}
S=-\left(\frac{\partial F}{\partial T}\right)_V
=k_B\beta^2\left(\frac{\partial F}{\partial\beta}\right)_V,
\end{equation}
we compute
\begin{align}
\left(\frac{\partial F}{\partial\beta}\right)_V
&=\frac{1}{\beta^2}\ln(2\cosh x)
-\frac{1}{\beta}\tanh x\left(\frac{\partial x}{\partial\beta}\right)_V,
\nonumber\\
&=\frac{1}{\beta^2}\left[\ln(2\cosh x)-x\tanh x\right].
\end{align}
Therefore
\begin{equation}
S(x)=k_B\left[\ln(2\cosh x)-x\tanh x\right].
\label{app:Sx}
\end{equation}

The same expression follows directly from the von Neumann entropy. Since the eigenvalues are $p_0=(1+r)/2$ and $p_1=(1-r)/2$,
\begin{equation}
S(r)=-k_B\left[
\frac{1+r}{2}\ln\left(\frac{1+r}{2}\right)
+\frac{1-r}{2}\ln\left(\frac{1-r}{2}\right)
\right].
\label{app:Sr}
\end{equation}
Using Eq.~\eqref{app:px},
\begin{align}
S
&=-k_B\left[
 p_0\ln\left(\frac{\ee^x}{2\cosh x}\right)
+p_1\ln\left(\frac{\ee^{-x}}{2\cosh x}\right)
\right],
\nonumber\\
&=k_B\left[\ln(2\cosh x)-x(p_0-p_1)\right],
\end{align}
which becomes Eq.~\eqref{app:Sx} because $p_0-p_1=\tanh x$.

\subsection{Generalized pressure and equation of state}

At fixed temperature (equivalently at fixed $\beta$), we have $\left(\frac{\partial x}{\partial V}\right)_\beta=-\frac{x}{V}$. The generalized pressure conjugate to $V$ is
\begin{equation}
P=-\left(\frac{\partial F}{\partial V}\right)_T.
\end{equation}
Differentiating Eq.~\eqref{app:F},
\begin{align}
\left(\frac{\partial F}{\partial V}\right)_\beta
&=-\frac{\Delta_0}{2V^2}
-\frac{1}{\beta}\tanh x\left(-\frac{x}{V}\right) =-\frac{\Delta_0}{2V^2}\left(1-\tanh x\right).
\end{align}
Thus $P(V,x)=\frac{\Delta_0}{2V^2}(1-\tanh x)$, and
\begin{equation}
P(V,r) = \frac{\Delta_0}{2V^2}(1-r).
\label{app:P}
\end{equation}
Comparison with Eq.~\eqref{app:U} yields the exact effective equation of state
\begin{equation}
U=PV.
\label{app:UPV}
\end{equation}
The effective equation of state presented in Eq.~\eqref{app:UPV} is a model-specific linear relationship derived from the spectral law $\Delta(V)\propto V^{-1}$. It should not be confused with the relationship $U=3PV/2$ for a three-dimensional monoatomic ideal gas. A useful dimensionless rewriting is
\begin{equation}
\frac{PV}{k_BT}=z(x),
\qquad
z(x):=x(1-\tanh x).
\label{app:z}
\end{equation}
The factor $z(x)$ in Eq.~\eqref{app:z} is analogous to a compressibility factor for real gases~\cite{MoranShapiro,CengelBoles}. Its derivative is
\begin{equation}
z'(x)=1-\tanh x-x\sech^2x.
\end{equation}
The unique positive maximum occurs at
\begin{equation}
1-\tanh x_\ast-x_\ast\sech^2x_\ast=0,
\qquad
x_\ast\simeq0.639232.
\label{app:xstar}
\end{equation}
This is a crossover scale of the chosen spectral law, not a thermodynamic critical point. Finally,
\begin{align}
F+TS
&=\frac{\Delta_0}{2V}
-\frac{1}{\beta}\ln(2\cosh x)
+\frac{1}{\beta}\left[\ln(2\cosh x)-x\tanh x\right]
\nonumber\\
&=\frac{\Delta_0}{2V}(1-\tanh x)=U,
\end{align}
which verifies the Legendre relation.

\section{First law, heat--work decomposition, and fundamental relation}
\label{app:firstlaw}

From $U=F+TS$ and $\dd F=-S\,\dd T-P\,\dd V$,
\begin{equation}
\dd U=T\,\dd S-P\,\dd V.
\label{app:firstlawthermo}
\end{equation}
With work performed by the system taken as positive,
\begin{equation}
\dd U=\delta Q-\delta W,
\qquad
\delta Q=T\,\dd S,
\qquad
\delta W=P\,\dd V.
\label{app:firstlawconv}
\end{equation}

\subsection{Microscopic decomposition}

Differentiating Eq.~\eqref{app:U} while treating $(V,r)$ as coordinates,
\begin{align}
\dd U
&=-\frac{\Delta_0}{2V^2}(1-r)\,\dd V
-\frac{\Delta_0}{2V}\,\dd r,
\nonumber\\
&=-P\,\dd V-\frac{\Delta(V)}{2}\,\dd r.
\label{app:dUvr}
\end{align}
Hence,
\begin{equation}
\delta W=P\,\dd V,
\qquad
\delta Q=-\frac{\Delta(V)}{2}\,\dd r.
\label{app:QW}
\end{equation}
The two terms have the standard microscopic interpretation: $\delta Q=\Tr(H\,\dd\rho)$ changes populations at fixed spectrum, while $-\delta W=\Tr(\rho\,\dd H)$ changes the spectrum at fixed populations. From Eq.~\eqref{app:Sr},
\begin{align}
\frac{\dd S}{\dd r}
&=-\frac{k_B}{2}\ln\left(\frac{1+r}{1-r}\right)
=-k_B\arctanh r
=-k_Bx.
\label{app:dSdr}
\end{align}
Therefore
\begin{equation}
T\,\dd S=-k_BTx\,\dd r
=-\frac{\Delta(V)}{2}\,\dd r
=\delta Q,
\end{equation}
where Eq.~\eqref{app:xdef} was used. This independently verifies the microscopic and thermodynamic identifications.

\subsection{Fundamental relation $S(U,V)$}

From $U = \frac{\Delta_0}{V}p_1$, let us define
\begin{equation}
y:=p_1=\frac{UV}{\Delta_0},
\qquad
p_0=1-y.
\label{app:y}
\end{equation}
Thus, the fundamental relation $S = S(U,V)$ is given by
\begin{equation}
S(U,V)=-k_B\left[(1-y)\ln(1-y)+y\ln y\right].
\label{app:fundamental}
\end{equation}
Substituting the expressions above, we obtain
\begin{align}
S(U,V)
&=
-k_B
\left[
\left(
1-\frac{UV}{\Delta_0}
\right)
\ln\left(
1-\frac{UV}{\Delta_0}
\right)\right.
+ \nonumber \\
&\left.+\frac{UV}{\Delta_0} 
\ln\left(
\frac{UV}{\Delta_0}
\right)
\right].
\end{align}
This is the fundamental relation for the volume-dependent qubit. It expresses the entropy as a state function of $U$ and $V$. The mathematical domain of the binary entropy is determined by $0\leq p_1\leq 1$, namely $0\leq UV/\Delta_0\leq 1$. However, throughout this work, we focus on positive-temperature Gibbs states, for which the lower-energy state is more populated than the excited state. Therefore, $p_1\leq 1/2$, and the physical domain becomes $0\leq \frac{UV}{\Delta_0}\leq\frac{1}{2}$, or, equivalently, $0\le y\le\frac12$, or $0\le U\le\frac{\Delta_0}{2V}=\frac{\Delta(V)}{2}=\frac{E_1(V)}{2}.$ Since
\begin{equation}
\frac{\dd S}{\dd y}=k_B\ln\frac{1-y}{y},
\end{equation}
we find
\begin{align}
\left(\frac{\partial S}{\partial U}\right)_V
&=k_B\frac{V}{\Delta_0}\ln\frac{1-y}{y},
\\
\left(\frac{\partial S}{\partial V}\right)_U
&=k_B\frac{U}{\Delta_0}\ln\frac{1-y}{y}.
\end{align}
For a Gibbs state, $(1-y)/y=p_0/p_1=\ee^{\beta\Delta(V)}$, hence
\begin{equation}
\ln\frac{1-y}{y}=\beta\Delta(V)=\frac{\beta\Delta_0}{V}.
\end{equation}
Therefore
\begin{equation}
\left(\frac{\partial S}{\partial U}\right)_V=k_B\beta=\frac1T,
\quad
\left(\frac{\partial S}{\partial V}\right)_U=k_B\beta\frac{U}{V}=\frac{P}{T},
\end{equation}
where Eq.~\eqref{app:UPV} was used in the second identity. Thus
\begin{equation}
\dd S=\frac{1}{T}\dd U+\frac{P}{T}\dd V,
\end{equation}
reproduces Eq.~\eqref{app:firstlawthermo}.

\subsection{Maxwell relations}

For the entropy representation, define
\begin{equation}
L(y):=\ln\frac{1-y}{y},
\qquad
L'(y)=-\frac{1}{y(1-y)}.
\end{equation}
Using
\begin{equation}
\frac1T=k_B\frac{V}{\Delta_0}L(y),
\qquad
\frac{P}{T}=k_B\frac{U}{\Delta_0}L(y),
\end{equation}
we obtain, at fixed $U$,
\begin{align}
\left(\frac{\partial(1/T)}{\partial V}\right)_U
&=\frac{k_B}{\Delta_0}L(y)
+k_B\frac{V}{\Delta_0}L'(y)\frac{U}{\Delta_0},
\nonumber\\
&=\frac{k_B}{\Delta_0}L(y)-\frac{k_B}{\Delta_0(1-y)}.
\end{align}
At fixed $V$,
\begin{align}
\left(\frac{\partial(P/T)}{\partial U}\right)_V
&=\frac{k_B}{\Delta_0}L(y)
+k_B\frac{U}{\Delta_0}L'(y)\frac{V}{\Delta_0},
\nonumber\\
&=\frac{k_B}{\Delta_0}L(y)-\frac{k_B}{\Delta_0(1-y)}.
\end{align}
Hence
\begin{equation}
\left(\frac{\partial(1/T)}{\partial V}\right)_U
=
\left(\frac{\partial(P/T)}{\partial U}\right)_V.
\label{app:maxwellS}
\end{equation}
The explicit calculations above show that the fundamental relation $S(U,V)$ generates the correct conjugate variables. Hence, the Maxwell relation in the entropy representation is satisfied identically. This also confirms that the pressure obtained from the entropy representation agrees with the pressure obtained from the Helmholtz free energy.

\subsubsection{Helmholtz representation}

Using $\dd U=T\,\dd S-P\,\dd V$ and taking the differential of $F = U-TS$, we obtain
\begin{equation}
dF
=
-S\,dT
-
P\,\dd V.
\end{equation}
To verify the Maxwell relation for this potential, note that
\begin{equation}
\frac{\dd S}{\dd x}=-k_Bx\sech^2x,
\qquad
\left(\frac{\partial x}{\partial V}\right)_T=-\frac{x}{V},
\end{equation}
so
\begin{equation}
\left(\frac{\partial S}{\partial V}\right)_T
=\frac{k_Bx^2}{V}\sech^2x
=\frac{\Delta_0^2}{4k_BT^2V^3}\sech^2x.
\end{equation}
Likewise,
\begin{equation}
\left(\frac{\partial x}{\partial T}\right)_V=-\frac{x}{T},
\end{equation}
and Eq.~\eqref{app:P} gives
\begin{equation}
\left(\frac{\partial P}{\partial T}\right)_V
=\frac{\Delta_0}{2V^2}\frac{x}{T}\sech^2x
=\frac{\Delta_0^2}{4k_BT^2V^3}\sech^2x.
\end{equation}
Thus,
\begin{equation}
\left(\frac{\partial S}{\partial V}\right)_T
=
\left(\frac{\partial P}{\partial T}\right)_V.
\label{app:maxwellF}
\end{equation}

\section{Response functions and information geometry}
\label{app:response}

\subsection{Heat capacity and energy fluctuations}

At fixed $V$,
\begin{equation}
\left(\frac{\partial x}{\partial T}\right)_V=-\frac{x}{T}.
\end{equation}
Using Eq.~\eqref{app:U},
\begin{align}
C_V
&=\left(\frac{\partial U}{\partial T}\right)_V
=\frac{\Delta_0}{2V}\left(-\sech^2x\right)\left(-\frac{x}{T}\right),
\nonumber\\
&=k_Bx^2\sech^2x.
\label{app:Cv}
\end{align}
The Hamiltonian satisfies $H^2=\Delta^2\Pi_1$. Hence
\begin{equation}
\Var_\rho(H)=\Delta^2p_1(1-p_1).
\end{equation}
Because
\begin{equation}
p_1(1-p_1)=\frac{1-r^2}{4}=\frac14\sech^2x,
\end{equation}
we obtain
\begin{equation}
\Var_\rho(H)=\frac{\Delta^2}{4}\sech^2x.
\end{equation}
Since $x=\beta\Delta/2$,
\begin{equation}
C_V=k_B\beta^2\Var_\rho(H).
\label{app:CvVar}
\end{equation}

For the commuting Gibbs family, the BKM covariance of $H$ reduces to the ordinary variance,
\begin{equation}
g_{\beta\beta}^{\mathrm{BKM}}=\Var_\rho(H),
\end{equation}
so Eq.~\eqref{app:CvVar} may be written
\begin{equation}
C_V=k_B\beta^2g_{\beta\beta}^{\mathrm{BKM}}.
\end{equation}

\subsection{Rank-one pullback metric}

With $\Xi:=\Pi_0-\Pi_1$, the same state can be written $\rho(x)=\ee^{x\Xi}/(2\cosh x)$. The one-parameter BKM metric is
\begin{equation}
\dd\ell^2=\Var_\rho(\Xi)\,\dd x^2=\sech^2x\,\dd x^2.
\label{app:lineelementx}
\end{equation}
Pulling this line element back to the $(\beta,V)$ control plane gives
\begin{equation}
\dd x=\frac{\Delta_0}{2V}\dd\beta
-\frac{\beta\Delta_0}{2V^2}\dd V,
\end{equation}
and therefore
\begin{equation}
\dd\ell^2
=\sech^2\!\left(\frac{\beta\Delta_0}{2V}\right)
\left(
\frac{\Delta_0}{2V}\dd\beta
-\frac{\beta\Delta_0}{2V^2}\dd V
\right)^2.
\label{app:pullback}
\end{equation}
The pullback has rank one. Its null directions satisfy $\dd x=0$, i.e.
\begin{equation}
\frac{\dd\beta}{\beta}=\frac{\dd V}{V},
\qquad
\frac{\beta}{V}=\mathrm{const.}
\end{equation}
These are exactly the finite-temperature isentropic leaves derived in Appendix~\ref{app:adiabats}. The density operator is unchanged, even though the Hamiltonian scale and the internal energy change.

\subsection{Isothermal compressibility and free-energy curvature}

The isothermal compressibility associated with the generalized coordinate is
\begin{equation}
\kappa_T=-\frac1V\left(\frac{\partial V}{\partial P}\right)_T
=-\frac1V\left[\left(\frac{\partial P}{\partial V}\right)_T\right]^{-1}.
\end{equation}
At fixed $T$, $(\partial x/\partial V)_T=-x/V$. Differentiating Eq.~\eqref{app:P},
\begin{align}
\left(\frac{\partial P}{\partial V}\right)_T
&=-\frac{\Delta_0}{V^3}(1-\tanh x)
+\frac{\Delta_0}{2V^2}\left[-\sech^2x\left(-\frac{x}{V}\right)\right],
\nonumber\\
&=-\frac{\Delta_0}{V^3}
\left[1-\tanh x-\frac{x}{2}\sech^2x\right].
\label{app:dPdV}
\end{align}
Define
\begin{equation}
\phi(x):=1-\tanh x-\frac{x}{2}\sech^2x.
\label{app:fx}
\end{equation}
Then
\begin{equation}
\kappa_T=\frac{V^2}{\Delta_0 \phi(x)},
\label{app:kappa}
\end{equation}
and
\begin{equation}
\left(\frac{\partial^2F}{\partial V^2}\right)_T
=\frac{\Delta_0}{V^3}\phi(x).
\label{app:Fcurv}
\end{equation}
The positive zero of $\phi$ is
\begin{equation}
x_\kappa\simeq1.10886.
\label{app:xkappa}
\end{equation}
Thus the generalized isothermal stiffness is positive for $0<x<x_\kappa$ and negative for $x>x_\kappa$. This sign change is a local property of the effective control coordinate and should not be interpreted as a many-body critical point.

\subsection{Thermal expansion coefficient}

At fixed pressure,
\begin{equation}
0=\dd P
=\left(\frac{\partial P}{\partial T}\right)_V\dd T
+\left(\frac{\partial P}{\partial V}\right)_T\dd V,
\end{equation}
so
\begin{equation}
\alpha_P:=\frac1V\left(\frac{\partial V}{\partial T}\right)_P
=-\frac1V
\frac{(\partial P/\partial T)_V}{(\partial P/\partial V)_T}.
\end{equation}
Using
\begin{equation}
\left(\frac{\partial P}{\partial T}\right)_V
=\frac{\Delta_0}{2V^2}\frac{x}{T}\sech^2x,
\end{equation}
and Eq.~\eqref{app:dPdV},
\begin{equation}
\alpha_P=\frac{x\sech^2x}{2T\phi(x)}.
\label{app:alpha}
\end{equation}
For $x>0$, the numerator is positive, so $\alpha_P$ and $\kappa_T$ have the same sign. In particular, $\alpha_P>0$ on the positive-stiffness branch $0<x<x_\kappa$.

\section{Reversible adiabats}
\label{app:adiabats}

Differentiating Eq.~\eqref{app:Sx},
\begin{equation}
\dd S=-k_Bx\sech^2x\,\dd x.
\label{app:dSdx}
\end{equation}
For finite positive temperature, $x>0$, so a reversible adiabat $\dd S=0$ satisfies
\begin{equation}
\dd x=0,
\qquad
x=\frac{\beta\Delta_0}{2V}=\mathrm{const.}
\label{app:adiabaticx}
\end{equation}
Equivalently,
\begin{equation}
TV=\mathrm{const.}
\label{app:TVconst}
\end{equation}
and $r=\tanh x$ is constant. Substituting constant $r$ into Eq.~\eqref{app:P},
\begin{equation}
PV^2=\frac{\Delta_0}{2}(1-r)=\mathrm{const.}
\label{app:PV2}
\end{equation}
Thus
\begin{equation}
PV^\gamma=\mathrm{const.},
\qquad
\gamma=2.
\end{equation}
Because $T\propto V^{-1}$ and $P\propto V^{-2}$ on the same curve,
\begin{equation}
\frac{P}{T^2}=\mathrm{const.}
\end{equation}
The value $\gamma=2$ follows from the spectral law $\Delta\propto V^{-1}$; it is not a phenomenological heat-capacity ratio.

\section{Detailed Carnot-cycle calculation}
\label{app:carnot}

Let $\beta_H<\beta_C$. The four corners are: $A=(\beta_H,V_A)$, $B=(\beta_H,V_B)$, $C=(\beta_C,V_C)$, and $D=(\beta_C,V_D).$ The adiabatic constraints $x_B=x_C$ and $x_D=x_A$ imply
\begin{equation}
V_C=\frac{\beta_C}{\beta_H}V_B,
\qquad
V_D=\frac{\beta_C}{\beta_H}V_A.
\label{app:carnotvolumes}
\end{equation}
The ordering $V_A<V_D<V_B<V_C$ additionally requires
\begin{equation}
\frac{V_B}{V_A}>\frac{\beta_C}{\beta_H}.
\end{equation}

\subsection{Generic isothermal formulas}

Along an isotherm $\beta=\beta_j$,
\begin{equation}
x=\frac{\beta_j\Delta_0}{2V},
\qquad
\dd x=-\frac{x}{V}\dd V,
\qquad
\dd r=\sech^2x\,\dd x.
\end{equation}
Equations~\eqref{app:P} and \eqref{app:QW} give
\begin{equation}
\delta W=-\frac1{\beta_j}(1-\tanh x)\,\dd x,
\qquad
\delta Q=-\frac{x}{\beta_j}\sech^2x\,\dd x.
\end{equation}
Hence, for an isothermal branch $i\to j$,
\begin{align}
W_{i\to j}
&=-\frac1{\beta_{\rm iso}}\int_{x_i}^{x_j}(1-\tanh x)\,\dd x,
\nonumber\\
&=\frac1{\beta_{\rm iso}}
\left[
 x_i-x_j+\ln\frac{\cosh x_j}{\cosh x_i}
\right],
\label{app:Wisoth}
\\
Q_{i\to j}
&=-\frac1{\beta_{\rm iso}}\int_{x_i}^{x_j}x\sech^2x\,\dd x,
\nonumber\\
&=\frac1{\beta_{\rm iso}}
\left[
 x_i\tanh x_i-x_j\tanh x_j
 +\ln\frac{\cosh x_j}{\cosh x_i}
\right].
\label{app:Qisoth}
\end{align}
We used here $\int x\sech^2x\,\dd x=x\tanh x-\ln(\cosh x) + \mathrm{const.}$

\subsection{Generic reversible-adiabatic formula}

Along an adiabat, $r=r_{\rm ad}$ and $Q=0$. Therefore
\begin{align}
\Delta U_{i\to j}
&=\frac{\Delta_0}{2}(1-r_{\rm ad})
\left(\frac1{V_j}-\frac1{V_i}\right),
\\
W_{i\to j}
&=-\Delta U_{i\to j}
=\frac{\Delta_0}{2}(1-r_{\rm ad})
\left(\frac1{V_i}-\frac1{V_j}\right).
\label{app:Wadiab}
\end{align}

\subsection{Four branches and their signs}

For the hot isothermal expansion $A\to B$, $x_A>x_B$. Equations~\eqref{app:Wisoth}--\eqref{app:Qisoth} give
\begin{align}
W_{AB}
&=\frac1{\beta_H}\left[
 x_A-x_B+\ln\frac{\cosh x_B}{\cosh x_A}
\right]>0,
\\
Q_{AB}
&=\frac1{\beta_H}\left[
 x_A\tanh x_A-x_B\tanh x_B
 +\ln\frac{\cosh x_B}{\cosh x_A}
\right]>0.
\end{align}
The signs follow immediately from their integral forms because $1-\tanh x>0$ and $x\sech^2x>0$ for $x>0$.

For the adiabatic expansion $B\to C$, $V_C>V_B$, $r_C=r_B$, and
\begin{equation}
Q_{BC}=0,
\qquad
W_{BC}=\frac{\Delta_0}{2}(1-r_B)
\left(\frac1{V_B}-\frac1{V_C}\right)>0.
\end{equation}

For the cold isothermal compression $C\to D$, $x_C=x_B$ and $x_D=x_A$, so
\begin{align}
W_{CD}
&=\frac1{\beta_C}\left[
 x_B-x_A+\ln\frac{\cosh x_A}{\cosh x_B}
\right]<0,
\\
Q_{CD}
&=\frac1{\beta_C}\left[
 x_B\tanh x_B-x_A\tanh x_A
 +\ln\frac{\cosh x_A}{\cosh x_B}
\right]<0.
\end{align}
Finally, for the adiabatic compression $D\to A$, $V_D>V_A$, $r_D=r_A$, and
\begin{equation}
Q_{DA}=0,
\qquad
W_{DA}=\frac{\Delta_0}{2}(1-r_A)
\left(\frac1{V_D}-\frac1{V_A}\right)<0.
\end{equation}
At fixed $\beta$, Eq.~\eqref{app:Ux} gives $U=\beta^{-1}z(x)$, where $z(x)$ is defined in Eq.~\eqref{app:z}. If the plotted cycle is restricted to $0<x_B<x_A<x_\ast$, then $z$ is increasing and
\begin{equation}
\Delta U_{AB}=\frac1{\beta_H}[z(x_B)-z(x_A)]<0,
\end{equation}
whereas
\begin{equation}
\Delta U_{CD}=\frac1{\beta_C}[z(x_A)-z(x_B)]>0.
\end{equation}
These additional sign statements require the monotonic branch of $z$; the Carnot efficiency itself does not.

\subsection{Net work and efficiency}

The two adiabats exchange no heat, so
\begin{equation}
W_{\rm cyc}=Q_{AB}+Q_{CD}.
\end{equation}
The bracket in $Q_{CD}$ is the negative of that in $Q_{AB}$, hence
\begin{equation}
Q_{CD}=-\frac{\beta_H}{\beta_C}Q_{AB}.
\end{equation}
Therefore
\begin{equation}
W_{\rm cyc}=\left(1-\frac{\beta_H}{\beta_C}\right)Q_{AB},
\end{equation}
and with $Q_{\rm in}=Q_{AB}$,
\begin{equation}
\eta_{\rm Carnot}=\frac{W_{\rm cyc}}{Q_{\rm in}}
=1-\frac{\beta_H}{\beta_C}
=1-\frac{T_C}{T_H}.
\label{app:carnoteff}
\end{equation}

\section{Detailed Otto-cycle calculation}
\label{app:otto}

The two volumes are
\begin{equation}
V_1=V_4=V_{\max},
\qquad
V_2=V_3=V_{\min},
\qquad
V_{\min}<V_{\max}.
\end{equation}
The reservoir temperatures are assigned to points $1$ and $3$:
\begin{equation}
1=(\beta_C,V_{\max}),
\qquad
3=(\beta_H,V_{\min}).
\end{equation}
The adiabatic conditions $x_1=x_2$ and $x_3=x_4$ give
\begin{equation}
\beta_2=\beta_C\frac{V_{\min}}{V_{\max}},
\qquad
\beta_4=\beta_H\frac{V_{\max}}{V_{\min}},
\label{app:betas24}
\end{equation}
and
\begin{equation}
r_1=r_2,
\qquad
r_3=r_4.
\end{equation}

\subsection{Isochoric and adiabatic branch formulas}

Along an isochore at fixed $V=V_{\rm iso}$, the work $\delta W=0$, so
\begin{equation}
\delta Q=\dd U=-\frac{\Delta_0}{2V_{\rm iso}}\,\dd r.
\end{equation}
Therefore,
\begin{equation}
Q_{i\to j}=\frac{\Delta_0}{2V_{\rm iso}}(r_i-r_j).
\label{app:Qisochore}
\end{equation}
The adiabatic work is given by Eq.~\eqref{app:Wadiab}.

\noindent For $1\to2$,
\begin{equation}
Q_{12}=0,
\qquad
W_{12}=\frac{\Delta_0}{2}(1-r_1)
\left(\frac1{V_{\max}}-\frac1{V_{\min}}\right)<0.
\end{equation}
For $2\to3$,
\begin{equation}
W_{23}=0,
\qquad
Q_{23}=\frac{\Delta_0}{2V_{\min}}(r_1-r_3).
\label{app:Q23}
\end{equation}
For $3\to4$,
\begin{equation}
Q_{34}=0,
\qquad
W_{34}=\frac{\Delta_0}{2}(1-r_3)
\left(\frac1{V_{\min}}-\frac1{V_{\max}}\right)>0.
\end{equation}
For $4\to1$,
\begin{equation}
W_{41}=0,
\qquad
Q_{41}=\frac{\Delta_0}{2V_{\max}}(r_3-r_1).
\label{app:Q41}
\end{equation}

\subsection{Engine condition, work, and efficiency}

Summing the two adiabatic work contributions,
\begin{equation}
W_{\rm Otto}
=\frac{\Delta_0}{2}(r_1-r_3)
\left(\frac1{V_{\min}}-\frac1{V_{\max}}\right).
\label{app:Wotto}
\end{equation}
Hence, positive work requires
\begin{equation}
r_1>r_3
\quad\Longleftrightarrow\quad
x_1>x_3
\quad\Longleftrightarrow\quad
\frac{V_{\max}}{V_{\min}}<\frac{\beta_C}{\beta_H}.
\label{app:ottoengine}
\end{equation}
In this regime,
\begin{equation}
Q_{\rm in}=Q_{23}=\frac{\Delta_0}{2V_{\min}}(r_1-r_3),
\end{equation}
and
\begin{equation}
Q_{\rm out}=-Q_{41}=\frac{\Delta_0}{2V_{\max}}(r_1-r_3).
\end{equation}
Therefore
\begin{equation}
\eta_{\rm Otto}
=\frac{W_{\rm Otto}}{Q_{\rm in}}
=1-\frac{V_{\min}}{V_{\max}}
=1-\frac{\epsilon_{\min}}{\epsilon_{\max}},
\label{app:ottoeff}
\end{equation}
where $\epsilon_{\min}=\Delta_0/V_{\max}$ and $\epsilon_{\max}=\Delta_0/V_{\min}$.
From Eq.~\eqref{app:ottoengine},
\begin{equation}
\frac{V_{\min}}{V_{\max}}>\frac{\beta_H}{\beta_C}=\frac{T_C}{T_H},
\end{equation}
so
\begin{equation}
\eta_{\rm Otto}<\eta_{\rm Carnot}
\end{equation}
for every nondegenerate engine. The functional form agrees with the classical Otto expression $\eta_{\mathrm{Otto}}^{\mathrm{class}} = 1-R^{1-\gamma}$, for $R = \frac{V_{\max}}{V_{\min}} > 1$, the volumetric compression ratio, when the effective adiabatic exponent $\gamma=2$ is inserted, but here $\gamma$ originates from the Hamiltonian control law rather than from a classical heat-capacity ratio.

\section{Thermalization mismatch and relative-entropy form of Otto irreversibility}
\label{app:mismatch}

In this section only, it is useful to use dimensionless entropy
\begin{equation}
s:=\frac{S}{k_B}=-\Tr(\rho\ln\rho).
\end{equation}
Then Eq.~\eqref{app:dSdr} becomes
\begin{equation}
\dd s=-x\,\dd r.
\label{app:dsdr}
\end{equation}
All entropy productions $\Sigma$ below are dimensionless; the corresponding physical entropy production is $k_B\Sigma$.

\subsection{Factorized Otto entropy production}

The reservoir entropy production over the Otto cycle is
\begin{equation}
\Sigma_{\rm Otto}=\beta_CQ_{\rm out}-\beta_HQ_{\rm in}.
\label{app:SigmaDef}
\end{equation}
Using the heat exchanges above,
\begin{align}
\Sigma_{\rm Otto}
&=\left[
\frac{\beta_C\Delta_0}{2V_{\max}}
-\frac{\beta_H\Delta_0}{2V_{\min}}
\right](r_1-r_3),
\nonumber\\
&=(x_1-x_3)(r_1-r_3).
\label{app:SigmaFactor}
\end{align}
In the engine regime, both factors are positive.

\subsection{Relative entropy between two Gibbs states on one isochore}

Introduce 
\begin{equation}
\Xi:=\Pi_0-\Pi_1,
\qquad
\rho(x)=\frac{\ee^{x\Xi}}{Z_x},
\end{equation}
with $Z_x=2\cosh x$. Then
\begin{align}
\ln\rho(x) &=x\Xi-\ln Z_x,\\
\Tr[\rho(x)\Xi]&=\frac{\partial}{\partial x}\ln Z_x=\tanh x=r(x).
\end{align}
For two members of the family,
\begin{align}
D(\rho(x_a)\Vert\rho(x_b))
&=\Tr\rho(x_a)\left[\ln\rho(x_a)-\ln\rho(x_b)\right],
\nonumber\\
&=(x_a-x_b)r_a+\ln\frac{\cosh x_b}{\cosh x_a}.
\label{app:Dgibbs}
\end{align}
The hot isochore maps $\rho_2=\rho(x_1)$ to $\rho_3=\rho(x_3)$, so
\begin{equation}
D(\rho_2\Vert\rho_3)
=(x_1-x_3)r_1+\ln\frac{\cosh x_3}{\cosh x_1}.
\end{equation}
The cold isochore maps $\rho_4=\rho(x_3)$ to $\rho_1=\rho(x_1)$, so
\begin{equation}
D(\rho_4\Vert\rho_1)
=(x_3-x_1)r_3+\ln\frac{\cosh x_1}{\cosh x_3}.
\end{equation}
Adding them cancels the logarithms:
\begin{equation}
D(\rho_2\Vert\rho_3)+D(\rho_4\Vert\rho_1)
=(x_1-x_3)(r_1-r_3).
\end{equation}
Combining with Eq.~\eqref{app:SigmaFactor},
\begin{equation}
\Sigma_{\rm Otto}
=D(\rho_2\Vert\rho_3)+D(\rho_4\Vert\rho_1).
\label{app:SigmaD}
\end{equation}

\subsection{General fixed-Hamiltonian mismatch identity}

Let $\tau_\beta:=\frac{\ee^{-\beta H}}{Z}$ be the Gibbs state selected by a reservoir during a fixed-Hamiltonian contact, and define
\begin{equation}
Q=\Tr[H(\rho_{\rm out}-\rho_{\rm in})],
\quad
\Delta s=s(\rho_{\rm out})-s(\rho_{\rm in}).
\end{equation}
Because $\ln\tau_\beta=-\beta H-\ln Z$,
\begin{equation}
D(\rho\Vert\tau_\beta)=-s(\rho)+\beta\Tr(H\rho)+\ln Z.
\end{equation}
Therefore
\begin{align}
D(\rho_{\rm in}\Vert\tau_\beta)
-D(\rho_{\rm out}\Vert\tau_\beta)
&=\Delta s-\beta Q.
\end{align}
Thus the algebraic identity is
\begin{equation}
\Sigma:=\Delta s-\beta Q
=D(\rho_{\rm in}\Vert\tau_\beta)
-D(\rho_{\rm out}\Vert\tau_\beta).
\label{app:generalMismatch}
\end{equation}
If the contact is generated by a Gibbs-preserving CPTP map $\Phi$ with $\Phi(\tau_\beta)=\tau_\beta$, monotonicity gives
\begin{equation}
D(\Phi(\rho_{\rm in})\Vert\tau_\beta)
\le D(\rho_{\rm in}\Vert\tau_\beta),
\end{equation}
so $\Sigma\ge0$. Under complete thermalization, $\rho_{\rm out}=\tau_\beta$, and
\begin{equation}
\Sigma=D(\rho_{\rm in}\Vert\tau_\beta).
\end{equation}

For a two-reservoir cyclic engine, let $Q_H>0$ and $Q_C=-|Q_C|<0$. Summing the dimensionless entropy production gives
\begin{equation}
\mathcal M_{\rm cyc}
=-\beta_HQ_H+\beta_C|Q_C|\ge0.
\end{equation}
Since $W_{\rm cyc}=Q_H-|Q_C|$,
\begin{equation}
W_{\rm cyc}
=\left(1-\frac{\beta_H}{\beta_C}\right)Q_H
-\frac{\mathcal M_{\rm cyc}}{\beta_C}.
\label{app:workMismatch}
\end{equation}
Hence
\begin{equation}
\eta
=\eta_{\rm Carnot}-\frac{\mathcal M_{\rm cyc}}{\beta_CQ_H}.
\label{app:etaMismatch}
\end{equation}
For the Otto cycle, $\mathcal M_{\rm cyc}=\Sigma_{\rm Otto}$, so
\begin{equation}
\eta_{\rm Otto}
=\eta_{\rm Carnot}
-\frac{D(\rho_2\Vert\rho_3)+D(\rho_4\Vert\rho_1)}{\beta_CQ_{\rm in}}.
\label{app:etaOttoD}
\end{equation}

\subsection{Finite-control origin and reversible limit}

Define the compression ratio
\begin{equation}
R:=\frac{V_{\max}}{V_{\min}}>1.
\end{equation}
Equation~\eqref{app:betas24} becomes
\begin{equation}
\beta_2=\frac{\beta_C}{R},
\qquad
\beta_4=\beta_HR.
\end{equation}
The mismatch entering Eq.~\eqref{app:SigmaFactor} can be written
\begin{align}
x_1-x_3
&=\frac{\Delta_0}{2V_{\min}}
\left(\frac{\beta_C}{R}-\beta_H\right),
\nonumber\\
&=\frac{\Delta_0}{2V_{\max}}
\left(\beta_C-\beta_HR\right).
\end{align}
Thus
\begin{equation}
\Sigma_{\rm Otto}
=\frac{\Delta_0}{2V_{\min}}
\left(\frac{\beta_C}{R}-\beta_H\right)(r_1-r_3).
\end{equation}
The reversible limit requires $\beta_2=\beta_H$ and $\beta_4=\beta_C$, i.e.
\begin{equation}
R\to\frac{\beta_C}{\beta_H}.
\end{equation}
Then $x_1\to x_3$, $r_1-r_3\to0$, and consequently
\begin{equation}
\Sigma_{\rm Otto}\to0,
\qquad
Q_{\rm in}\to0,
\qquad
W_{\rm Otto}\to0.
\end{equation}
Thus the Otto efficiency approaches the Carnot value only in the degenerate zero-work limit.

\section{Unitary-generated nonpassivity as a nonequilibrium battery}
\label{app:nonpassivity}

This section derives the main-text result concerning a work resource prepared by a unitary at fixed Hamiltonian. The central distinction is between an apparent heat-based efficiency, which ignores the charging work, and the complete work accounting, which includes it.

\subsection{Maximum energy-basis coherence on the Gibbs unitary orbit}

At a fixed preparation point $(\beta_{\rm np},V_{\rm np})$, define
\begin{equation}
H_{\rm np}=\Delta_{\rm np}\Pi_1,
\quad
\Delta_{\rm np}=\frac{\Delta_0}{V_{\rm np}},
\quad
r_{\rm np}=\tanh\!\left(\frac{\beta_{\rm np}\Delta_{\rm np}}{2}\right).
\end{equation}
In the energy basis, write an arbitrary state on the unitary orbit of the Gibbs state as
\begin{equation}
\rho=\begin{pmatrix}
\frac12+\zeta & c\\
c^* & \frac12-\zeta
\end{pmatrix}.
\end{equation}
Unitary equivalence fixes the determinant:
\begin{equation}
\frac14-\zeta^2-|c|^2=p_0p_1=\frac{1-r_{\rm np}^2}{4}.
\end{equation}
Hence
\begin{equation}
|c|^2=\frac{r_{\rm np}^2}{4}-\zeta^2.
\end{equation}
The off-diagonal magnitude, and therefore the $l_1$-coherence $C_{l_1}=2|c|$, is maximal for $\zeta=0$:
\begin{equation}
|c|_{\max}=\frac{r_{\rm np}}{2}.
\end{equation}
A representative maximizer is
\begin{equation}
\rho_{\rm coh}^{\max}
=\frac12\begin{pmatrix}
1 & \,r_{\rm np}\\
\,r_{\rm np} & 1
\end{pmatrix}.
\label{app:rhocoh}
\end{equation}
The unitary that creates this state changes both coherence and energy-basis populations; the thermodynamic resource is therefore its induced nonpassivity, not coherence by itself.

\subsection{Charging cost and ergotropy}

The equilibrium energy is
\begin{equation}
U_{\rm eq}=\Delta_{\rm np}p_1
=\frac{\Delta_{\rm np}}{2}(1-r_{\rm np}),
\end{equation}
whereas Eq.~\eqref{app:rhocoh} has excited-state population $1/2$, so
\begin{equation}
U_{\rm coh}^{\max}=\frac{\Delta_{\rm np}}{2}.
\end{equation}
The isolated charging unitary therefore stores
\begin{equation}
\Delta F_{\rm np}=\Delta U_{\rm np}
=U_{\rm coh}^{\max}-U_{\rm eq}
=\frac{\Delta_{\rm np}}{2}r_{\rm np}.
\label{app:DeltaFnp}
\end{equation}
The unitary operation does not change the entropy, so the nonequilibrium free-energy increase equals the energy increase. The original Gibbs state is the passive rearrangement of the same spectrum. Consequently, the ergotropy of $\rho_{\rm coh}^{\max}$ is
\begin{align}
\mathcal W_{\max}
&=\Tr[H_{\rm np}(\rho_{\rm coh}^{\max}-\rho_{\rm eq})]
=\Delta F_{\rm np},\nonumber \\
&=\frac{\Delta_{\rm np}}{2}
\tanh\!\left(\frac{\beta_{\rm np}\Delta_{\rm np}}{2}\right).
\label{app:ergotropy}
\end{align}
Defining $x_{\rm np}:=\frac{\beta_{\rm np}\Delta_{\rm np}}{2}$, we have
\begin{equation}
\beta_{\rm np}\Delta F_{\rm np}=x_{\rm np}\tanh x_{\rm np}.
\label{app:betadF}
\end{equation}

\subsection{Closed auxiliary charge--discharge loop}

To compare an assisted cycle with the same thermal cycle, the auxiliary unitary protocol must close before the next reservoir contact. Let
\begin{equation}
\rho_{\rm coh}^{\max}=\mathcal V_{\rm ch}\rho_{\rm eq}\mathcal V_{\rm ch}^\dagger,
\qquad
\rho_{\rm eq}=\mathcal V_{\rm dis}\rho_{\rm coh}^{\max}\mathcal V_{\rm dis}^\dagger.
\end{equation}
The auxiliary loop is isolated and performed at fixed Hamiltonian:
\begin{equation}
Q_{\rm aux}=0,
\qquad
H_{\rm before}=H_{\rm after}=H_{\rm np}.
\end{equation}
It also returns the boundary state exactly to the unassisted value. Therefore, the subsequent hot heat input and thermalization mismatch are unchanged.

Let the unassisted thermal cycle satisfy Eq.~\eqref{app:workMismatch},
\begin{equation}
W_{\rm cyc}=\eta_{\rm Carnot}Q_H-\frac{\mathcal M_{\rm cyc}}{\beta_C}.
\end{equation}
If the battery is optimally discharged but its charging cost is deliberately omitted from the engine bookkeeping, the apparent work output is
\begin{equation}
W_{\rm app}^{\max}=W_{\rm cyc}+\Delta F_{\rm np},
\end{equation}
and the apparent heat-based efficiency is
\begin{equation}
\eta_{\rm th,app}^{\max}
=\frac{W_{\rm cyc}+\Delta F_{\rm np}}{Q_H}
=\eta_{\rm Carnot}
+\frac{\beta_C\Delta F_{\rm np}-\mathcal M_{\rm cyc}}{\beta_CQ_H}.
\label{app:etaapp}
\end{equation}
Therefore
\begin{equation}
\eta_{\rm th,app}^{\max}>\eta_{\rm Carnot}
\quad\Longleftrightarrow\quad
\beta_C\Delta F_{\rm np}>\mathcal M_{\rm cyc}.
\label{app:appcond}
\end{equation}
This is not a violation of the Carnot bound because $\Delta F_{\rm np}$ was supplied as external work. The charging work cost is exactly
\begin{equation}
W_{\rm prep}^{\rm cost}=\Delta F_{\rm np}.
\end{equation}
After this cost is debited,
\begin{equation}
W_{\rm net}^{\max}
=W_{\rm app}^{\max}-W_{\rm prep}^{\rm cost}
=W_{\rm cyc},
\end{equation}
so
\begin{equation}
\eta_{\rm net}^{\max}
=\frac{W_{\rm net}^{\max}}{Q_H}
=\eta_{\rm Carnot}-\frac{\mathcal M_{\rm cyc}}{\beta_CQ_H}
\le\eta_{\rm Carnot}.
\label{app:etanet}
\end{equation}

If the preparation point is at the cold inverse temperature, $\beta_{\rm np}=\beta_C$, then Eq.~\eqref{app:betadF} gives
\begin{equation}
\beta_C\Delta F_{\rm np}=x_C\tanh x_C,
\qquad
x_C:=\frac{\beta_C\Delta_0}{2V_{\rm np}}.
\end{equation}
The apparent threshold becomes
\begin{equation}
x_C\tanh x_C>\mathcal M_{\rm cyc}.
\end{equation}
Since $\dd(x\tanh x)/\dd x=\tanh x+x\sech^2x>0$ for $x>0$, this threshold has a unique positive solution. For $\mathcal M_{\rm cyc}\ll1$,
\begin{equation}
x\tanh x=x^2-\frac{x^4}{3}+O(x^6),
\end{equation}
so the leading threshold is
\begin{equation}
x_C\gtrsim\sqrt{\mathcal M_{\rm cyc}}.
\end{equation}

\subsection{Partial discharge and residual mismatch}

Let $\mathcal V_d$ be an arbitrary unitary discharge,
\begin{equation}
\rho_d=\mathcal V_d\rho_{\rm coh}^{\max}\mathcal V_d^\dagger,
\end{equation}
with extracted work
\begin{equation}
W_d=\Tr[H_{\rm np}(\rho_{\rm coh}^{\max}-\rho_d)].
\end{equation}
Because $\rho_d$, $\rho_{\rm coh}^{\max}$, and $\rho_{\rm eq}$ are isospectral,
\begin{equation}
s(\rho_d)=s(\rho_{\rm eq}).
\end{equation}
Using $\ln\rho_{\rm eq}=-\beta_{\rm np}H_{\rm np}-\ln Z$,
\begin{equation}
D(\rho_d\Vert\rho_{\rm eq})
=\beta_{\rm np}\Tr[H_{\rm np}(\rho_d-\rho_{\rm eq})].
\end{equation}
But
\begin{align}
\Tr[H_{\rm np}(\rho_d-\rho_{\rm eq})]
&=\Tr[H_{\rm np}(\rho_{\rm coh}^{\max}-\rho_{\rm eq})]
-W_d
\nonumber\\
&=\mathcal W_{\max}-W_d.
\end{align}
Therefore, a complete thermal reset after the partial discharge produces the residual mismatch
\begin{equation}
\Sigma_{\rm res}
=D(\rho_d\Vert\rho_{\rm eq})
=\beta_{\rm np}(\mathcal W_{\max}-W_d).
\label{app:residual}
\end{equation}
Equivalently,
\begin{equation}
\beta_{\rm np}W_d+\Sigma_{\rm res}
=\beta_{\rm np}\mathcal W_{\max}
=x_{\rm np}\tanh x_{\rm np}.
\label{app:balancebattery}
\end{equation}
The two endpoints are transparent: $W_d=0$ converts the entire stored work value into thermalization mismatch, while $W_d=\mathcal W_{\max}$ returns the system to its passive Gibbs representative and leaves no residual mismatch.

For an explicit finite reservoir $R$, the reduced-system mismatch is only one part of the information balance. For a product reference $\rho_{\rm eq}\otimes\rho_E$,
\begin{equation}
D(\rho'_{SE}\Vert\rho_{\rm eq}\otimes\rho_E)
=D(\rho'_S\Vert\rho_{\rm eq})
+D(\rho'_E\Vert\rho_E)
+I(S{:}E)_{\rho'},
\label{app:chainrule}
\end{equation}
therefore, reservoir perturbations and system--reservoir correlations provide additional nonnegative contributions to the full bookkeeping.

\end{document}